\def\ifundefined#1{\expandafter\ifx\csname#1\endcsname\relax}

\ifundefined{arXiv}
    \documentclass[journal]{IEEEtran}
\else
    \documentclass[10pt]{article}
\fi

\usepackage{xcolor} 
\ifundefined{arxiv}
\else
    \usepackage{IEEEtrantools}

    \newcommand\blfootnote[1]{%
        \begingroup
        \renewcommand\thefootnote{}\footnote{#1}%
        \addtocounter{footnote}{-1}%
        \endgroup
    }
\fi

\usepackage{cite} 

\usepackage[pdftex]{graphicx}
\graphicspath{{../figures/}}
\DeclareGraphicsExtensions{.pdf} 

\usepackage{tikz}       
\usepackage{pgfplots}   
\pgfplotsset{compat=1.15} 

\usepackage{amsmath}
\usepackage{amsthm}
\usepackage{amsfonts}
\usepackage{amssymb}

\usepackage{algorithmic}

\usepackage{array} 

\ifundefined{arxiv}
\else
    \usepackage{accents}
    \usepackage{geometry}[margin = 1in]
    \usepackage{setspace}
\fi

\usepackage[caption=false,font=footnotesize]{subfig}

\usepackage{stfloats} 

\usepackage{url} 

\makeatletter
    \@ifpackageloaded{xcolor}{}{\usepackage{xcolor}}
\makeatother
\newcommand{\uppercaseGreek}[1]{%
    \begingroup
    \ucmathlist\MakeUppercase{#1}%
    \endgroup
}

\newcommand{\ucmathlist}{%
    \def\alpha{\mathrm{A}}%
    \def\beta{\mathrm{B}}%
    \let\gamma=\Gamma
    \let\delta=\Delta
    \def\epsilon{\mathrm{E}}%
    \def\varepsilon{\mathrm{E}}%
    \def\zeta{\mathrm{Z}}%
    \def\eta{\mathrm{H}}%
    \let\theta=\Theta
    \let\vartheta=\Theta
    \def\iota{\mathrm{I}}%
    \def\kappa{\mathrm{K}}%
    \let\lambda=\Lambda
    \def\mu{\mathrm{M}}%
    \def\nu{\mathrm{N}}%
    \let\xi=\Xi
    \let\pi=\Pi
    \let\varpi=\Pi
    \def\rho{\mathrm{P}}%
    \def\varrho{\mathrm{P}}%
    \let\sigma=\Sigma
    \def\tau{\mathrm{T}}%
    \let\upsilon=\Upsilon
    \let\phi=\Phi
    \let\varphi=\Phi
    \def\chi{\mathrm{X}}%
    \let\psi=\Psi
    \let\omega=\Omega
}

\makeatletter
    \@ifpackageloaded{amsthm}{}{\usepackage{amsthm}}
\makeatother
\theoremstyle{plain}
\theoremstyle{definition}

\makeatletter
\def\renewtheorem#1{%
    \expandafter\let\csname#1\endcsname\relax
    \expandafter\let\csname c@#1\endcsname\relax
    \gdef\renewtheorem@envname{#1}
    \renewtheorem@secpar
}
\def\renewtheorem@secpar{\@ifnextchar[{\renewtheorem@numberedlike}{\renewtheorem@nonumberedlike}}
\def\renewtheorem@numberedlike[#1]#2{\newtheorem{\renewtheorem@envname}[#1]{#2}}
\def\renewtheorem@nonumberedlike#1{
    \def\renewtheorem@caption{#1}
    \edef\renewtheorem@nowithin{\noexpand\newtheorem{\renewtheorem@envname}{\renewtheorem@caption}}
    \renewtheorem@thirdpar
}
\def\renewtheorem@thirdpar{\@ifnextchar[{\renewtheorem@within}{\renewtheorem@nowithin}}
\def\renewtheorem@within[#1]{\renewtheorem@nowithin[#1]}
\makeatother
\input{auxFiles/symbolDef.sty}
\input{auxFiles/riceColors.sty}
\def\myfactor{0.2}
\input{auxFiles/auxTikzDefs.sty}
\usepackage{tabularx}

\def\Keff{\ensuremath{\schK^{\text{eff}}}}
\def\Kthres{\ensuremath{\schK^{\text{thres}}}}
\def\NN{\mathsf{NN}}
\def\RNN{\mathsf{RNN}}
\def\GNN{\mathsf{GNN}}

\begin{document}



\def\theTitle{Unsupervised Learning of Sampling Distributions for Particle Filters}

\title{\theTitle}


\ifundefined{arXiv}
    \author{Fer\hspace{0.015cm}nando~Gama,~%
            Nicolas~Zilberstein,~%
            Martin~Sevilla,~%
            Richard~G.~Baraniuk,~%
            and~Santiago~Segarra
    \thanks{This work was partially supported by the NSF under award CCF-2008555. F. Gama was with the Department of Electrical and Computer Engineering, Rice University, Houston, TX 77005 USA. N. Zilberstein, M. Sevilla, R. Baraniuk, and S. Segarra are with the Department of Electrical and Computer Engineering, Rice University, Houston, TX 77005 USA. Email: fgama@ieee.org, \{nzilberstein,msevilla,richb,segarra\}@rice.edu. Partial results have appeared in \cite{Gama2022-UnrollingParticles}.}}
\else
    \author{Fer\hspace{0.015cm}nando Gama, Nicolas Zilberstein, Martin Sevilla, Richard G. Baraniuk, and Santiago Segarra}
    \date{}
\fi


\ifundefined{arxiv}
    \markboth{IEEE TRANSACTIONS ON SIGNAL PROCESSING (SUBMITTED)}%
             {Gama \MakeLowercase{\textit{et al.}}: \theTitle}
\else
\fi


\maketitle


\begin{abstract}
    Accurate estimation of the states of a nonlinear dynamical system is crucial for their design, synthesis, and analysis.
    Particle filters are estimators constructed by simulating trajectories from a sampling distribution and averaging them based on their importance weight.
    For particle filters to be computationally tractable, it must be feasible to simulate the trajectories by drawing from the sampling distribution.
    Simultaneously, these trajectories need to reflect the reality of the nonlinear dynamical system so that the resulting estimators are accurate.
    Thus, the crux of particle filters lies in designing sampling distributions that are both easy to sample from and lead to accurate estimators.
    In this work, we propose to \emph{learn} the sampling distributions.
    We put forward four methods for learning sampling distributions from observed measurements.
    Three of the methods are parametric methods in which we learn the mean and covariance matrix of a multivariate Gaussian distribution; each methods exploits a different aspect of the data (generic, time structure, graph structure). 
    The fourth method is a nonparametric alternative in which we directly learn a transform of a uniform random variable.
    All four methods are trained in an unsupervised manner by maximizing the likelihood that the states may have produced the observed measurements.
    Our computational experiments demonstrate that learned sampling distributions exhibit better performance than designed, minimum-degeneracy sampling distributions.
    %
\end{abstract}


\ifundefined{arxiv}
    %
    \IEEEpeerreviewmaketitle
\else
    \blfootnote{This work is supported by grants....}
    \blfootnote{The authors are with the Department of Electrical and Computer Engineering, Rice University, Houston, TX 77005 USA. Email: \{fgama,nzilberstein,msevilla,richb,segarra\}@rice.edu}}
    \blfootnote{Partial results have appeared in \cite{Gama2022-UnrollingParticles}.}
\fi


\section{Introduction} \label{sec:intro}


Nonlinear dynamical systems serve as models for a wide range of problems in science and engineering. 
For example, due to the natural hysteresis of the material, nonlinear dynamical systems are used to describe electrical circuits involving ferromagnetic inductors \cite{Kumar2018-Ferromagnetic}. 
Mostly, they have been very popular tools in control theory \cite{Khalil2015-NonlinearControl}. They play a key role in designing and synthesizing controllers for spacecraft systems \cite{Mehrjardi2018-Spacecraft}, in managing energy consumption of electrical vehicles \cite{Alamdari2020-EV}, in reducing ripple in wind power systems \cite{Mahmodicherati2019-Wind}, and even in real-time bidding for programmatic advertising \cite{Karlsson2020-Advertising}.

Due to their practical relevance, research on nonlinear dynamical systems has a long history \cite{Isidori1995-NonlinearControl}. 
Topics such as the existence and uniqueness of solutions, dependence on initial conditions, stability of the systems, as well as perturbation analysis have dominated the field \cite{Khalil2002-NonlinearSystems}.
Recent results concern dissipativity and its connection to stability \cite{Hill2022-Dissipativity}, identification \cite{Batselier2022-Identification}, and oscillations \cite{Koshkin2022-Oscillation}.

In this work, we focus on estimating the states of the system \cite{Simon2006-OptimalStateEstimation, Sawo2009-NonlinearEstimation}. 
Almost any decision to be made in the synthesis, design, or analysis of nonlinear dynamical systems needs to rely on a solid knowledge of the system state.
An inaccurate modeling of the system or an impossibility to access the state --and instead being able to measure some function of the state-- are the major hindrances in the task of estimation.

Many estimators have been developed, making different assumptions on the system to reach different levels of accuracy guarantees.
For instance, assuming linear dynamics with Gaussian noise leads to the linear least-squares estimator \cite{Kailath2000-LinearEstimation}.
Another example is assuming the model is nonlinear but follows the Markov property on the conditionality of its transition probabilities \cite{AndersonMoore1979-OptimalFiltering}.
In this scenario, oftentimes the maximum \textit{a posteriori} estimate can be obtained.
These approaches, however, either oversimplify the model or can become computationally intractable due to the high-dimensional integrals involved.

Particle filtering has risen as an algorithmic tool that is capable of estimating the state in a computationally efficient way \cite{Doucet2000-ParticleFiltering, Djuric2003-ParticleFiltering, Godsill2019-ParticleFiltering}.
In essence, particle filtering consists of simulating plausible trajectories of the system of interest, and then carrying out a weighted average of these trajectories to obtain accurate estimators \cite{Bugallo2017-AdaptiveIS}.
The challenge in particle filtering is in designing a sampling distribution that is simultaneously good at generating realistic trajectories and easy to sample from \cite{Elvira2021-AdvancesParticleFiltering, Branchini2021-OptimalParticleFiltering}.

Particle filters leverage the law of large numbers to provide certain guarantees on the accuracy of the estimators and convergence to the true value of the state if enough particles are simulated.
However, these results oftentimes rely on using specific sampling distributions that may only be computationally efficient in limited scenarios \cite{Elvira2017-AdaptingNumParticles}.
Furthermore, particle filters suffer from weight degeneracy, a phenomenon that causes only a few of the simulated trajectories to be meaningful, severely impacting the accuracy of the resulting estimator \cite{Elvira2022-EffectiveSampleSize}.
Thus, most of the research on particle filtering has revolved around the appropriate design of sampling distributions \cite{Elvira2018-ImprovedAuxiliaryParticleFilters, Elvira2019-ElucidatingAuxiliary}.
In this work, however, we leverage modern deep neural network techniques to learn the sampling distribution from observed measurements --without access to the true trajectories.

Deep neural networks consist of a cascade of blocks, each of which applies a linear transformation followed by an activation function that is, typically, nonlinear \cite{Murphy2012-ProbabilisticML}.
The blocks are known as layers, and the number of these layers in cascade determines the depth of the neural networks.
The exact matrix values to be used in the linear transforms at each layer are typically determined by a gradient-descent algorithm (or a variant thereof) in an attempt to minimize some loss function over the observed data.
The process of determining the actual value of the linear transforms is known as training \cite{Goodfellow2016-DeepLearning}.

In this work, we propose four different deep neural network frameworks that are used to learn the sampling distribution of particle filters, three of which are parametric --learning the mean and covariance matrix of a Gaussian distribution-- and one which is non-parametric --learning an arbitrary transform of a uniform random variable.
These frameworks lead, naturally, to distributions that are easy to sample from (either multivariate normal or uniform distributions).
We train these deep neural networks in an unsupervised manner.
This means that we only need access to a trajectory of measurements, but not to the true value of the states.
By learning the deep neural network parameters that maximize the likelihood of the observed trajectories under the given model, we are able to obtain a sampling distribution that is capable of simulating good trajectories.
In short, learning the sampling distribution is, generally, easier than designing it, while it also allows for more flexibility and adaptability.

Many works have proposed using learning to improve particle filtering. Typically, learning is used mostly to estimate the model or transform the variables into spaces more amenable for sampling \cite{Brock2018-ParticleDifferentiable}. 
Recently, conditional normalizing flows were proposed as parameterization of the proposal distribution~\cite{chen2021differentiable}.
Recurrent neural networks (RNNs) have also been used to learn the mean of a multivariate normal sampling distribution as well as the particle weights \cite{Karkus2020-ParticleRNN}.
Most importantly, in all these cases, the neural networks are trained using supervised learning. 
This requires access to true trajectories of the system, which are typically unavailable, and it does not guarantee that the trajectories observed at test time will be similar enough to ensure generalization. 
We address this fundamental drawback by proposing a trainable particle filter based on unsupervised learning.

In Section~\ref{sec:PF} we review the basics of particle filtering and introduce the notation.
In Section~\ref{sec:unsupervised} we present the unsupervised learning framework of sampling distributions.
\begin{itemize}
    \item  First, we assume the sampling distribution is a multivariate normal and we learn a time-dependent mean and covariance matrix using a fully-connected neural network (Sec.~\ref{subsec:MLP}).
    \item Second, we consider a recurrent neural network (RNN) architecture for learning the mean and covariance matrix (Sec.~\ref{subsec:RNN}). 
    RNNs are good at keeping track of past values of the trajectory, potentially allowing the distribution to learn from samples that are located further in the past.
    \item Third, we consider architectures that exploit the data structure. In particular, we consider graph neural networks (GNNs) which are architectures tailored to process graph-based data (Sec.~\ref{subsec:GNN}). 
    GNNs can be particularly useful when dealing with large distributed nonlinear dynamical systems, where the components have sparse connections between them.
    \item Fourth, we consider a non-parametric approach (Sec.~\ref{subsec:arbitrary}).
    More specifically, we sample from a uniform distribution, and we use a deep neural network to learn an arbitrary nonlinear transform between the uniform distribution and a random variable that represents the trajectory of states.
\end{itemize}
We close Section~\ref{sec:unsupervised} by explaining the details of unsupervised learning through maximization of the likelihood of the model (Sec.~\ref{subsec:model}).
In Section~\ref{sec:sims}, we run a series of simulated examples to showcase the performance of learned sampling distributions as opposed to designed baselines.
We consider linear Gaussian (Sec.~\ref{subsec:linear}), nonlinear Gaussian (Sec.~\ref{subsec:nonlinear}), linear  non-Gaussian (Sec.~\ref{subsec:nongaussian}), and nonlinear non-Gaussian (Sec.~\ref{subsec:sir}) dynamical systems. 
In general, we observe that learned sampling distributions exhibit better performance than designed, minimum-degeneracy sampling distributions \cite{Doucet2000-ParticleFiltering}.
Finally, we draw conclusions in Section~\ref{sec:conclusions}.


\section{Particle Filtering} \label{sec:PF}

Consider a dynamical system described by a sequence of states $\{\vcx_{t}\}_{t \geq 0}$ with $\vcx_{t} \in \fdR^{N}$ for all $t \in \fdN_{0}$.
The transition between states in different time instants is considered to satisfy the Markov property and, thus, is completely characterized by the transition distribution given by
\begin{equation} \label{eq:transition}
    \vcx_{t}|\vcx_{t-1} \distro p(\vcx_{t}|\vcx_{t-1})
\end{equation}
for all $t \geq 1$.
The initial state is distributed following
\begin{equation} \label{eq:initialState}
    \vcx_{0} \distro p(\vcx_{0}).
\end{equation}
These states are considered unobservable.
Instead, there is a sequence of measurements $\{\vcy_{t}\}_{t \geq 0}$, with $\vcy_{t} \in \fdR^{M}$ for all $t \in \fdN_{0}$, that is accessible.
Given the current value of the state, the measurements are distributed as follows
\begin{equation} \label{eq:measurement}
    \vcy_{t}|\vcx_{t} \distro p(\vcy_{t}|\vcx_{t})
\end{equation}
for all $t \geq 0$.
These three distributions \eqref{eq:transition}--\eqref{eq:measurement} are considered known.

The objective is to estimate a target quantity $\vcz_{t}$ that depends on the true states of the system $\vcx_{0:t} = \{\vcx_{0},\ldots,\vcx_{t}\}$.
We denote this as $\vcz_{t} = \fnf_{t}(\vcx_{0:t})$ for some (possibly time-varying) mapping $\fnf_{t}$.
To estimate $\vcz_{t}$ from a sequence of observations $\vcy_{0:t} = \{\vcy_{0},\ldots,\vcy_{t}\}$, we use the conditional expectation to construct an estimator $\vctz_{t}$ as
\begin{equation} \label{eq:conditionalExpectation}
    \vctz_{t} = \xp\big[ \vcz_{t} | \vcy_{0:t} \big] = \intinfty \fnf_{t}(\vcx_{0:t}) \fnp(\vcx_{0:t}|\vcy_{0:t})\ d\vcx_{0:t}.
\end{equation}
Using Bayes' rule, the posterior distribution of the state trajectory given the measurements $\fnp(\vcx_{0:t}|\vcy_{0:t})$ can be written recursively as
\begin{equation} \label{eq:recursiveBayes}
    \fnp(\vcx_{0:t}|\vcy_{0:t}) = \fnp(\vcx_{0:t-1}|\vcy_{0:t-1}) \frac{\fnp(\vcy_{t}|\vcx_{t}) \fnp(\vcx_{t}|\vcx_{t-1})}{\fnp(\vcy_{t}|\vcy_{0:t-1})}.
\end{equation}
Note that the numerator of the update rule, i.e. $\fnp(\vcy_{t}|\vcx_{t})\fnp(\vcx_{t}|\vcx_{t-1})$, can be computed directly from \eqref{eq:transition}--\eqref{eq:measurement}.
The denominator $\fnp(\vcy_{t}|\vcy_{0:t-1})$ can also be computed from \eqref{eq:transition}--\eqref{eq:measurement} by marginalizing over all possible states at times $\vcx_{t}$ and $\vcx_{t-1}$ and using the previous step in the recursion $\vcp(\vcx_{0:t-1}|\vcy_{0:t-1})$. Then, since all the distributions are known, it should be technically possible to compute the conditional expectation estimator \eqref{eq:conditionalExpectation}.
However, this entails high-dimensional integrals with are typically intractable.
Therefore, \eqref{eq:conditionalExpectation} cannot usually be used in practice.

The law of large numbers can be leveraged to suggest a practical estimator consisting of taking $K$ samples $\{\vcx_{0:t}^{(k)}\}_{k=1}^{K}$, independently, identically distributed as $\vcx_{0:t}^{(k)} \distro \fnp(\vcx_{0:t}|\vcy_{0:t})$, and then averaging them.
Note, however, that even if we had access to the posterior $\fnp(\vcx_{0:t}|\vcy_{0:t})$, it may still be intractable to sample from it.

Particle filtering consists of sampling $\{\vcx_{0:t}^{(k)}\}_{k}$ from some other distribution $\vcx_{0:t}^{(k)} \distro \fnpi(\vcx_{0:t}|\vcy_{0:t})$ and computing the estimate as
\begin{equation} \label{eq:particleFiltering}
    \vchz_{t} = \sum_{k=1}^{K} \schw_{t}^{(k)} \fnf_{t}\big(\vcx_{0:t}^{(k)} \big) \ , \ \schw_{t}^{(k)} = \frac{\sctw_{t}^{(k)}}{\sum_{k'=1}^{K} \sctw_{t}^{(k')}},
\end{equation}
where the normalized weights $\schw_{t}^{(k)}$ are computed from the set of unnormalized weights $\{\sctw_{t}^{(k)}\}_{k}$, each of which is given by $\sctw_{t}^{(k)} = \fnp(\vcy_{0:t}|\vcx_{0:t}^{(k)}) \fnp(\vcx_{0:t}^{(k)}) / \fnpi(\vcx_{0:t}^{(k)}|\vcy_{0:t})$.
For the estimate in \eqref{eq:particleFiltering} to be tractable, sampling from $\fnpi(\vcx_{0:t}|\vcy_{0:t})$ has to be computationally feasible.
If no further restrictions are imposed on the sampling distribution $\fnpi(\vcx_{0:t}|\vcy_{0:t})$, then the particle filtering method receives the name of Bayesian Importance Sampling, the distribution $\fnpi$ is known as the importance function, and the weights are known as the importance weights \cite{Geweke1989-Econometrics}. The samples $\{\vcx_{0:t}^{(k)}\}_{k}$ are often referred to as particles or trajectories.

To further facilitate computational tractability in particle filtering, the sampling distribution $\fnpi$ is typically restricted to have the form
\begin{equation} \label{eq:samplingDistribution}
    \fnpi(\vcx_{0:t}|\vcy_{0:t}) = \fnpi(\vcx_{0}|\vcy_{0}) \prod_{\tau=1}^{t} \fnpi(\vcx_{\tau}|\vcx_{0:\tau-1}, \vcy_{0:\tau}).
\end{equation}
This implies that $\fnpi(\vcx_{0:t}|\vcy_{0:t})$ can be computed recursively over time.
Then, for each time $t$, it suffices to sample $\vcx_{t}^{(k)} \distro \fnpi(\vcx_{t}|\vcx_{0:t-1}^{(k)}, \vcy_{0:t})$.
The unnormalized weights can be computed recursively as well, following
\begin{equation} \label{eq:unnormalizedWeights}
    \sctw_{t} ^{(k)} = \sctw_{t-1}^{(k)} \frac{\fnp(\vcy_{t}|\vcx_{t}^{(k)}) \fnp(\vcx_{t}^{(k)}|\vcx_{t-1}^{(k)})}{\fnpi(\vcx_{t}^{(k)}|\vcx_{0:t-1}^{(k)},\vcy_{0:t})}.
\end{equation}
Particle filtering with sampling distributions of the form \eqref{eq:samplingDistribution} is often known as Sequential Importance Sampling \cite{Bugallo2017-AdaptiveIS}.

While computationally convenient, adopting a sampling distribution as in \eqref{eq:samplingDistribution} causes the particle filtering to suffer from weight degeneracy.
This means that the unconditional variance of the weights, considering the observations $\vcy_{0:t}$ as random variables, increases over time \cite{Kong1994-Degeneracy}.
The practical implications are that, over time, only one particle carries all the weight while the rest become insignificant.
This affects the quality of the estimator \eqref{eq:particleFiltering} as it virtually relies on a single trajectory.
The variance of the weights can be minimized, conditional upon $\vcx_{0:t-1}^{(k)}$ and $\vcy_{0:t}$, if the sampling distribution is chosen to be such that \cite{Chen1996-Predictive}
\begin{equation} \label{eq:minDeg}
    \fnpi(\vcx_{t}|\vcx_{0:t-1}^{(k)}, \vcy_{0:t}) = \fnp(\vcx_{t}|\vcx_{t-1}^{(k)}, \vcy_{t}).
\end{equation}
In this case, the unnormalized weight updates become $\sctw_{t}^{(k)} = \sctw_{t-1}^{(k)} \fnp(\vcy_{t}|\vcx_{t-1}^{(k)})$.
Sampling from $\fnp(\vcx_{t}|\vcx_{t-1}^{(k)},\vcy_{t})$, however, is generally intractable.

Weight degeneracy can be minimized [cf. \eqref{eq:minDeg}] but it cannot be avoided completely.
Thus, resampling is typically used to reduce its impact on the estimator \eqref{eq:particleFiltering}.
In short, resampling consists of randomly sampling trajectories according to their weight, thus giving more importance to the trajectories that carry larger weights.
More specifically, the level of weight degeneracy is measured by
\begin{equation} \label{eq:Keff}
    \Keff_{t} = \frac{1}{\sum_{k=1}^{K} (\schw_{t}^{(k)})^{2}},
\end{equation}
which is a proxy for the number of particles that can be considered to be effectively contributing to the estimator.
If this number of effective particles drop below a certain threshold $\Kthres$, then the trajectories are resampled following a distribution that assigns a probability $\sctw_{t}^{(k)}$ of choosing trajectory $k$.
After sampling $K$ times, a new set of $K$ trajectories is obtained (some of them likely to be repeated), and the weights are reset to be $1/K$ for all particles.
It is evident that using resampling affects the i.i.d. assumption on the particles, and thus many theoretical results, such as convergence, no longer hold  \cite{Liu1996-Metropolized}.

The particle filter is a computationally simple estimator for nonlinear systems that has shown significant success.
For it to yield good results, however, the sampling distribution $\fnpi$ has to be carefully designed in such a way that it is both easy to sample from and leads to good estimators \eqref{eq:particleFiltering}.
Many design methods have been proposed \cite{Elvira2021-AdvancesParticleFiltering}.
In what follows, instead of designing it, we propose and discuss several architectures to learn the sampling distribution from data, in an unsupervised manner.

\section{Unsupervised Learning of Sampling Distributions} \label{sec:unsupervised}

Designing a good sampling distribution $\fnpi$ for the particle filter that is simultaneously easy to sample from and yields acceptable performance is a challenging problem.
In what follows, we propose to use the sequence of measurements $\{\vcy_{t}\}_{t \geq 0}$ to \emph{learn} a suitable sampling distribution.
First, we parametrize the sampling distribution with a multivariate normal and use algorithm unrolling to learn the mean and covariance matrix from data (Section~\ref{subsec:MLP}).
Second, we use a recurrent neural network (RNN) that learns a hidden state that keeps track of past values of the trajectory (Section~\ref{subsec:RNN}).
Third, we use a graph neural network that exploits the data structure of the data improving the scalabality of the model (Section~\ref{subsec:GNN}).
Fourth, we learn an arbitrary transform comprised of multi-layer perceptrons (Section~\ref{subsec:arbitrary}).
Finally, we discuss how to learn these sampling distributions in an unsupervised manner using only the sequence of available measurements (Section~\ref{subsec:model}).
Simulation results in a myriad of different scenarios can be found in Section~\ref{sec:sims}.


\subsection{Multivariate normal parametrization} \label{subsec:MLP}

One distribution that is easy to sample from is the multivariate normal distribution.
Therefore, we choose to use this distribution to parametrize the sampling distribution $\fnpi$.
The multivariate normal distribution is completely characterized by a mean vector and a covariance matrix.
Since it is necessary for the sampling distribution $\fnpi$ to depend on the trajectory $\{\vcx_{\tau}^{(k)}\}_{\tau=0}^{t-1}$ and the measurements $\{\vcy_{\tau}\}_{\tau=0}^{t}$ up to the current time $t$, we propose to learn a mapping between these and the mean and covariance matrix of the multivariate normal distribution.

In particular, we consider fully-connected neural networks (also known as multi-layer perceptrons; MLPs) that, inspired by \eqref{eq:minDeg}, take as input the previous state $\vcx_{t-1}^{(k)}$ and the current measurement $\vcy_{t}$ and return the mean and covariance matrix of the multivariate normal distribution. Namely,
\begin{equation}
    \vcx_{t}^{(k)} \distro \fnpi(\vcx_{t}^{(k)}|\vcx_{t-1}^{(k)}, \vcy_{t}) = \gaussDistro \big( \vcmu_{t}, \mtSigma_{t} \big),
\end{equation}
with mean vector given by
\begin{equation} \label{eq:meanLearn}
    \vcmu_{t} = \vcmu_{t}\big( \vcx_{t-1}^{(k)}, \vcy_{t} \big)
\end{equation}
and covariance matrix given by
\begin{equation} \label{eq:covLearn}
    \mtSigma_{t} = \mtSigma \big(\vcx_{t-1}^{(k)}, \vcy_{t} \big).
\end{equation}
The equalities in both \eqref{eq:meanLearn} and \eqref{eq:covLearn} are used to represent that the mean vector and covariance matrix actually depend on the previous value of the state $\vcx_{t-1}^{(k)}$ for the $k$th trajectory and on the measurement $\vcy_{t}$ at time $t$.

The mapping between $\vcx_{t-1}^{(k)}$  and $\vcy_{t}$ and the mean $\vcmu_{t}$ at time $t$ is given by a fully-connected neural network $\NN_{t}^{\mu}$
\begin{equation}
    \vcmu_{t}\big( \vcx_{t-1}^{(k)}, \vcy_{t} \big)  = \NN_{t}^{\mu} \big( \vcx_{t-1}^{(k)}, \vcy_{t} \big).
\end{equation}
This is a cascade of blocks, each of which applies an affine transform characterized by matrix $\mtW_{t,\ell}^{\mu}$ and offset vector $\vcb_{t, \ell}^{\mu}$ for block $\ell$, followed by an activation function $\fnrho: \fdR \to \fdR$ that is applied element-wise to the output of the affine transform \cite[Ch. 6]{Goodfellow2016-DeepLearning}.
This can be compactly written as
\begin{equation} \label{eq:meanNN}
    \NN_{t}^{\mu} \big( \vcx_{t-1}^{(k)}, \vcy_{t} \big) = \vcz_{t,L}^{\mu} \text{ with } \vcz_{t, \ell}^{\mu} = \fnrho \big( \mtW_{t, \ell}^{\mu} \vcz_{t, \ell-1}^{\mu} + \vcb_{t, \ell}^{\mu} \big)
\end{equation}
for $L$ blocks, so that $\ell=1,\ldots,L$.
Essentially, each block (the affine transform followed by the activation function) is applied to the output of the previous block, forming a cascade.
The input $\vcz_{t,0}^{\mu}$ to the first block is given by the concatenation of the previous state and the current measurement $[(\vcx_{t-1}^{(k)})^{\Tr}, \vcy_{t}^{\Tr}]^{\Tr} \in \fdR^{N + M}$.
This implies that the matrix in the first affine transform $\mtW_{t, 1}^{\mu}$ is of size $F_{1} \times (M+N)$ where the value of $F_{1}$ is known as the number of (hidden) features at the output of block $1$.
We also have that $\vcb_{t,1}^{\mu} \in \fdR^{F_{1}}$.
In general, $\mtW_{t,\ell}^{\mu} \in \fdR^{F_{\ell} \times F_{\ell-1}}$ and $\vcb_{\ell} \in \fdR^{F_{\ell}}$, so that each block transforms the $F_{\ell-1}$ input features into $F_{\ell}$ output features.
The output of the multi-layer perceptron is the output of the last layer $\vcz_{t,L}^{\mu} \in \fdR^{F_{L}}$.
Note that, since this output represents the value of the mean 
$\vcz_{t,L}^{\mu} = \vcmu_{t}$, it has to hold that $F_{L} = N$ which is the size of the sampled state $\vcx_{t}^{(k)}$.
The activation function $\fnrho:\fdR \to \fdR$ is applied elementwise to the output of each affine transform, and therefore does not alter the dimensions.

In machine learning, the set of matrices and vectors that form the affine transforms of each block $\stTheta_{t}^{\mu} = \{\mtW_{t, \ell}^{\mu}, \vcb_{t, \ell}^{\mu}\}_{\ell=1}^{L}$ are called the parameters of the fully-connected neural network.
These are typically learned from data by solving an optimization problem \cite[Ch. 8]{Goodfellow2016-DeepLearning}.
See Section~\ref{subsec:model} for more details.
The number of blocks $L$, and the number of features $F_{\ell}$ at the output of each block $\ell=1,\ldots,L-1$ are design choices and are known as hyperparameters.
While there exist methods for choosing hyperparameters \cite{Bergstra11-Hyperparameter}, they are typically determined by experimentation.
The activation function $\fnrho$ is also a design choice and is typically a nonlinear function such as a rectified linear unit $\fnrho(x) = \ReLU(x) = \max\{x, 0\}$ or a hyperbolic tangent $\fnrho(x) = \tanh(x)$.

We remark that, in \eqref{eq:meanNN}, we are choosing to model the mapping from the previous state value $\vcx_{t-1}^{(k)}$ and the measurement $\vcy_{t}$ to the target mean value $\vcmu_{t}$ by means of a different fully-connected neural network for each time instant, as indicated by the subscript $t$.
This approach is known as the unrolling of the algorithm \cite{Monga2021-AlgorithmUnrolling}.

There are two main reasons for choosing a fully-connected neural network to parametrize the mapping from the previous state value $\vcx_{t-1}^{(k)}$ and the current measurement $\vcy_{t}$ to the target mean value $\vcmu_{t} = \NN_{t}^{\mu}(\vcx_{t-1}^{(k)}, \vcy_{t})$.
From a theoretical perspective, fully-connected neural networks can approximate any Borel measurable function with an arbitrary degree of accuracy if the number of features is large enough.
This is known as the universal approximation theorem \cite{Hornik89-Universal, Barron94-Bounds}.
From a practical perspective, neural networks are somewhat easy to train with gradient-based methods due to the fact that their learnable parameters are in the linear operation of the architecture, and not in the nonlinear one.
This makes the optimization problems easier to solve \cite[Ch. 8]{Goodfellow2016-DeepLearning}

To map the previous state and the current measurement to the covariance matrix, we propose to first leverage a fully-connected neural network to learn a representation of the data, and then build a distance-based kernel from it. More specifically,
\begin{equation} \label{eq:covMatrix}
    \mtSigma_{t} = \mtC \ \fnK(\vcz_{t}) \ \mtC^{\Tr}
\end{equation}
where $\mtC \in \fdR^{N\times N}$ is a matrix that is learnable from data, and where $\fnK(\vcz_{t}) \in \fdR^{N \times N}$ is a distance-based kernel matrix such as
\begin{equation} \label{eq:covKernel}
    [\fnK(\vcz_{t})]_{ij} = \exp \big( - ([\vcz_{t}]_{i} - [\vcz_{t}]_{j})^{2} \big)
\end{equation}
for the representation $\vcz_{t}$ obtained from a neural network as
\begin{equation} \label{eq:covNN}
    \vcz_{t} = \vcz(\vcx_{t-1}^{(k)}, \vcy_{t}) = \NN^{\Sigma}(\vcx_{t-1}^{(k)}, \vcy_{t}).
\end{equation}
Basically, we first transform the previous state $\vcx_{t-1}^{(k)}$ and the current measurement $\vcy_{t}$ into a representation vector $\vcz_{t} \in \fdR^{N}$ as in~\eqref{eq:covNN}. Then we compute the distance-based kernel matrix $\fnK(\vcz_{t}) \in \fdR^{N \times N}$ as in~\eqref{eq:covKernel}. Finally, we learn matrix $\mtC \in \fdR^{N \times N}$ to account for possible changes in direction and rotations of the variance components as specified in~\eqref{eq:covMatrix}.
Overall, the resulting covariance matrix $\mtSigma_{t}$ is guaranteed to be non-negative definite.

Note that the covariance matrix $\mtSigma_{t}$ is different for each time instant because the input to the neural network \eqref{eq:covNN} used to build the representation $\vcz_{t}$ changes with time.
However, the learning architectures are the same for all time instants.
The set of parameters to learn from data is comprised of the matrices $\mtW_{\ell}^{\Sigma} \in \fdR^{F_{\ell} \times F_{\ell-1}}$ and offset vectors $\vcb_{\ell}^{\Sigma} \in \fdR^{F_{\ell}}$ of each layer of the neural network in \eqref{eq:covNN}, as well as the matrix $\mtC$ in \eqref{eq:covMatrix}.
Thus, the set of learnable parameters $\stTheta^{\Sigma} = \{ \{\mtW_{\ell}^{\Sigma}, \vcb_{\ell}^{\Sigma}\}_{\ell=1}^{L}, \mtC\}$ is the same for all time instants.
For the sake of completeness, we note that, similar to \eqref{eq:meanNN}, the input to the neural network \eqref{eq:covNN} is the concatenation of the previous state and the current measurement $\vcz_{t,0}^{\Sigma} = [(\vcx_{t-1}^{(k)})^{\Tr}, \vcy_{t}^{\Tr}]^{\Tr}$ so that $F_{0} = N+M$.
Likewise, the representation in \eqref{eq:covNN} is collected as the output of the last layer $\vcz_{t} = \vcz_{t,L}^{\Sigma}$ so that $F_{L} = N$.
The decision to make the architecture that learns the covariance matrix fixed with time is to avoid the number of parameters growing proportionally to both time and the square of the dimension of the state (through the learnable matrix $\mtC$). 
This is different from the architecture for learning the mean, in that the latter which grows proportionally to time and the dimension of the state -- not quadratically with it.


\subsection{Recurrent neural networks} \label{subsec:RNN}

Parametrizing the sampling distribution $\fnpi$ with a multivariate normal distribution makes it easy to sample from.
Learning the mean and covariance matrix as described in Sec.~\ref{subsec:MLP} only takes into account the current measurement and the immediate past value of the state.
While this is suggested by the minimum-degeneracy sampling distribution \eqref{eq:minDeg}, it may be the case that including past information beyond the immediate one is helpful.
To do so in a way that avoids ever-increasing dimensionality, we consider recurrent neural networks (RNNs) \cite[Ch. 10]{Goodfellow2016-DeepLearning}, \cite{Graves2012-RNN}.

RNNs are machine learning architectures conceived to learn from sequential data.
Given an input sequence, they learn an internal representation known as a hidden state.
This hidden state is expected to capture past information from the sequence that is relevant for the task at hand.
More specifically, given the input sequence, which in our case is $\{[(\vcx_{t-1}^{(k)})^{\Tr}, \vcy_{t}^{\Tr}]^{\Tr}\}_{t}$, a sequence of internal states $\{\vcz_{t}\}_{t}$ for $\vcz_{t} \in \fdR^{H}$ is computed as
\begin{equation} \label{eq:hiddenRNN}
    \vcz_{t} = \fnrho \Big(\mtA \vcz_{t-1} + \mtB \begin{bmatrix}\vcx_{t-1}^{(k)} \\ \vcy_{t}\end{bmatrix} \Big).
\end{equation}
The matrices $\mtA \in \fdR^{H \times H}$ and $\mtB \in \fdR^{H \times (M+N)}$ are the parameters learned from data.
Note that these matrices are the same for all time instants, and thus, unlike the method in Sec.~\ref{subsec:MLP}, the number of parameters to learn does not depend on the length of the sequence.

The mean of the multivariate normal can then be learned at each time instant by computing an affine transform on the hidden state
\begin{equation} \label{eq:meanRNN}
    \vcmu_{t} = \mtW_{\RNN}^{\mu} \vcz_{t}+ \vcb_{\RNN}^{\mu},
\end{equation}
with $\mtW_{\RNN}^{\mu} \in \fdR^{N \times H}$ and $\vcb_{\RNN}^{\mu} \in \fdR^{N}$ being learnable parameters.
For the covariance matrix, we compute the kernel of an affine transform of the hidden state [cf. \eqref{eq:covMatrix}]
\begin{equation} \label{eq:covRNN}
    \mtSigma_{t} = \mtC\ \fnK(\mtW_{\RNN}^{\Sigma} \vcz_{t}+ \vcb_{\RNN}^{\Sigma})\ \mtC^{\Tr},
\end{equation}
where $\mtW_{\RNN}^{\Sigma} \in \fdR^{N \times H}$, $\vcb_{\RNN}^{\Sigma} \in \fdR^{N}$ and $\mtC \in \fdR^{N \times N}$ are all learnable parameters.

The set of learnable parameters for the RNN is $\stTheta^{\RNN} = \{ \mtA, \mtB, \mtC, \mtW_{\RNN}^{\mu}, \vcb_{\RNN}^{\mu}, \mtW_{\RNN}^{\Sigma}, \vcb_{\RNN}^{\Sigma}\}$.
Note that this set is independent of time, which allows for scalability to arbitrarily long sequences.
The information about past states and measurements is captured by the sequence of hidden states $\{\vcz_{t}\}$.
Thus, learning the mean and covariance by means of \eqref{eq:hiddenRNN}--\eqref{eq:covRNN} has the potential to leverage information that is not directly accessible to the method proposed in Sec.~\ref{subsec:MLP}, albeit in compressed form.

The computation of the hidden state as in \eqref{eq:hiddenRNN} may be unable to capture long-term dependencies \cite{Bengio1994-LongTermDependencies, Pascanu2013-TrainingRNN}.
To overcome this, we actually employ long short-term memory (LSTM) architectures.
These architectures add a series of gating strategies to be able to control the influx of present information with respect to past values of the input sequence.
We note, however, that since LSTMs are a particular implementation of RNNs, the computation of the hidden state is conceptually similar to \eqref{eq:hiddenRNN}.
The details on the specific gating mechanisms of LSTMs can be found in \cite{Hochreiter1997-LSTM, Gers2000-LSTM}.
Another alternative architecture includes gated recurrent units (GRUs) and its description can be found in \cite{Bengio2015-GRU, Jozefowicz2015-GRU}.
Finally, we would like to note that the expressivity of the hidden state can be enhanced by considering deep RNNs, instead of single-layer ones.
The same holds for the mean vector and the covariance matrix, where the affine transforms in \eqref{eq:meanRNN} and \eqref{eq:covRNN} can be replaced by fully-connected neural networks.


\subsection{Exploiting data structure: Graph neural networks} \label{subsec:GNN}

The fully-connected NN architecture used to map the previous state and the current measurement to the mean vector and covariance matrix of a multivariate normal (Sec.~\ref{subsec:MLP}) learns an affine transform from $(N+M)$-dimensional vectors to $N$-dimensional ones.
The RNN architecture (Sec.~\ref{subsec:RNN}) maps analogous dimensions but does so in a way that exploits the time structure by learning a hidden state that keeps track of past values of the measurement sequence and the simulated trajectory.
The number of learnable parameters in both architectures depends on the size of the state $N$ and on the size of the measurements $M$.
Oftentimes, the states or measurements may present additional structure that can further regularize the number of learnable parameters to be independent of either $N$ or $M$, improving scalability.

One such particular case is that of graph neural networks (GNNs) \cite{Gama2018-Archit, Gama2020-GNNs, Ruiz2021-GNNs}.
GNNs consider the input to be a graph signal \cite{Sandryhaila2013-DSPG, Shuman2013-SPG, Sandryhaila2014-DSPGfreq}.
Given a graph support $\stG = (\stV, \stE)$ where $\stV$ is a set of $N$ nodes and $\stE \subseteq \stV \times \stV$ is the set of edges, a graph signal associates an $F$-dimensional vector $\vcx_{n} \in \fdR^{F}$ to each node, $n=1,\ldots,N$.
The collection of $N$ vectors of dimension $F$ can be compactly written in a matrix $\mtX \in \fdR^{N\times F}$ where each row is given by the vector $\vcx_{n}^{\Tr}$.
Graph signals can be used to model measurements that arise in distributed plants \cite{Gama2022-DistributedLQR}, power grids \cite{Owerko2022-Power}, communication networks \cite{Chowdhury2021-UWMMSE}, brain activity~\cite{Medaglia1373}, teams of autonomous agents~\cite{Gama2022-ControlGNN}, among many others \cite{Bronstein2017-GeometricDL}.

To exploit the underlying graph structure, we require an operation that only relates measurements if they are connected by an edge.
Towards this end, let $\mtS \in \fdR^{N\times N}$ be a matrix description of the graph, i.e. it satisfies that $[\mtS]_{ij} = 0$ whenever $(j,i) \notin \stE$ for $i \neq j$.
The most popular choices of graph matrix descriptions in the literature include the adjacency matrix \cite{Sandryhaila2013-DSPG, Sandryhaila2014-DSPGfreq}, the Laplacian matrix \cite{Shuman2013-SPG, Bronstein2017-GeometricDL}, the random walk matrix \cite{Heimowitz2017-MarkovGSP}, and their normalized counterparts.
Then, we can define the graph convolutional filter \cite{Segarra2017-GraphFilterDesign} as a linear operation on the input graph signal $\mtX$ whose output is computed by means of a $D$-order polynomial on the graph matrix description $\mtS$
\begin{equation} \label{eq:graphFilter}
    \mtY = \fnW(\mtX; \mtS) = \sum_{d=0}^{D} \mtS^{d} \mtX \mtW_{d}.
\end{equation}
The operation $\mtS^{d}\mtX$ in \eqref{eq:graphFilter} gathers information located at the $d$-hop neighborhood of each node by means of $d$ successive exchanges with one-hop neighbors.
Multiplying $\mtS^{d} \mtX$ by the filter coefficients in $\mtW_{d} \in \fdR^{F \times G}$ determines how much weight to assign to the information at each $d$-hop neighborhood.
Note that the output is a graph signal $\mtY \in \fdR^{N \times G}$ consisting of $G$-dimensional vectors at each node, $\vcy_{n} \in \fdR^{G}$.
The set of $(D+1)$ filter coefficients $\{\mtW_{k}\}_{d=0}^{D}$ amounts to $FG(D+1)$ parameters which may be learned from data.

A GNN is a particular case of the fully-connected neural network, where the affine transform is replaced by a graph convolutional filter \eqref{eq:graphFilter}.
Then, the mean vector can be learned as $\vcmu_{t} = \vcmu_{t}(\vcx_{t-1}^{(k)}, \vcy_{t}) = \GNN_{t}^{\mu}(\vcx_{t-1}^{(k)}, \vcy_{t}; \mtS)=\mtZ_{t, L}^{\mu}$ where
\begin{equation} \label{eq:meanGNN}
\mtZ_{t,\ell}^{\mu} = \fnrho \big( \fnW_{t,\ell}^{\mu}(\mtZ_{t,\ell-1}^{\mu}; \mtS) + \mtB_{t,\ell}^{\mu} \big).
\end{equation}
Here, the graph convolutional filter at each layer $\fnW_{t,\ell}^{\mu}$ is of order $D_{t,\ell}$ and maps $F_{t,\ell-1}$-dimensional input graph signals into $F_{t,\ell}$-dimensional output graph signals.
Thus, each graph filter is characterized by a set of $(D_{t,\ell}+1)$ filter coefficients $\{\mtW_{t,\ell,d}^{\mu}\}_{d=0}^{D_{t,\ell}}$, where $\mtW_{t,\ell,d}^{\mu} \in \fdR^{F_{t,\ell-1} \times F_{t,\ell}}$.
The offset matrix $\mtB_{t,\ell} \in \fdR^{N \times F_{t,\ell}}$ is actually computed as $\mtB_{t,\ell} = [b_{t,\ell,1} \vcOnes_{N}, \ldots, b_{t,\ell,F_{t,\ell}} \vcOnes_{N}]$ where the $f$th column is $b_{t,\ell,f} \vcOnes_{N}$ with $\vcOnes_{N} \in \fdR^{N}$ a vector of all ones and $b_{t,\ell,f}$ the learnable coefficient, for $f=1,\ldots,F_{t,\ell}$.
The input to the GNN is given by $\mtZ_{t,0}^{\mu} = [\vcx_{t-1}^{(k)}, \mtA_{t}^{\mu} \vcy_{t}] \in \fdR^{N \times 2}$ so that $F_{0} = 2$.
The matrix $\mtA_{t}^{\mu} \in \fdR^{N \times M}$ is used to adapt the potentially different dimensions of the measurement and the state.
The number of features at the output of the GNN is $F_{t,L}=1$ so that $\mtZ_{t,L}^{\mu} \in \fdR^{N\times F_{t,L}}$ becomes an $N$-dimensional vector that is used as the mean vector for the multivariate normal distribution $\vcz_{t,L}^{\mu} = \vcmu_{t}$.
Finally, we note that the set of learnable parameters for the GNN-based architecture for learning the mean vector is given by $\stTheta_{t}^{\mu,\GNN} = \{ \{\mtW_{t,\ell,d}^{\mu}\}_{d=0}^{D_{t,\ell}}, \{b_{t,\ell,f}^{\mu}\}_{f=1}^{F_{t,\ell}} \}_{\ell=1}^{L}$.
This amounts, for each time $t$, to $\sum_{\ell=1}^{L}(F_{t,\ell}F_{t,\ell-1}(D_{t,\ell}+1)+F_{t,\ell})$ learnable parameters, a quantity determined by design choices and independent of the size of the measurement $N$.
Thus, the dimensionality of the optimization landscape is also independent of $N$, allowing for scalability.

To learn the covariance matrix, we follow the same scheme as in \eqref{eq:covMatrix}--\eqref{eq:covNN}, except that we replace \eqref{eq:covNN} with a GNN.
The input to the first layer, then, is analogous to that of the GNN-based architecture for learning the mean vector, except it may potentially have a different adaptation matrix $\mtA^{\Sigma} \in \fdR^{M \times N}$, i.e. $\mtZ_{t,0}^{\Sigma} = [\vcx_{t-1}^{(k)}, \mtA^{\Sigma} \vcy_{t}] \in \fdR^{N \times 2}$ so that $F_{0} = 2$.
The output $\mtZ_{t,L}^{\Sigma} \in \fdR^{N \times F_{L}}$ is set to be a vector $\vcz_{t,L}^{\Sigma}$, meaning $F_{L} = 1$, and then is fed into \eqref{eq:covKernel} and later into \eqref{eq:covMatrix}.
The GNN is independent of time in the same way that the fully-connected NN in \eqref{eq:covNN} is.
This makes the set of learnable parameters to be $\stTheta^{\Sigma,\GNN} = \{ \{\mtW_{\ell,d}^{\mu}\}_{d=0}^{D_{\ell}}, \{b_{\ell,f}^{\mu}\}_{f=1}^{F_{\ell}} \}_{\ell=1}^{L}$.
Note, however, that since the matrix $\mtC$ in \eqref{eq:covMatrix} is of size $N \times N$, the overall number of learnable parameters required for the covariance matrix does depend on $N$.
Additionally, the matrices $\{\{\mtA_{t}^{\mu}\}, \mtA^{\Sigma}\}$ can either be learned or designed to reflect some sort of topological structure on the measurements as well.
Note, however, that if they are learned, then the number of learnable parameters will depend on $N$, potentially hindering scalability.

GNNs exploit the assumption that the measurements exhibit a graph-based structure.
They are naturally distributed architectures, meaning that each node in the graph can compute its output separately, requiring only to communicate with nearby neighbors.
They are also better at generalizing for graph-based data, since they are permutation equivariant and stable to deformations of the graph support \cite{Gama2020-Stability, Roddenberry2023-Local}.
There is a vast body of literature on GNNs, whereby graph convolutional filters \eqref{eq:graphFilter} can be replaced by non-convolutional filters \cite{Isufi2022-EdgeNets} or where the pointwise activation functions can be replaced by graph-based activation functions \cite{Ruiz2019-Nonlinear}.
Also, graph RNNs have been developed to handle sequences of graph signals \cite{Ruiz2020-GRNN}.
The key conceptual aspect to highlight is that, if the measurements present certain data structures, it could be convenient to consider learning architectures that exploit such structures \cite[Ch. 9]{Goodfellow2016-DeepLearning}.


\subsection{Arbitrary transform} \label{subsec:arbitrary}

All the learning methods discussed in the previous sections parameterize the sampling distribution as a multivariate normal.
Then, each of them provides different ways of learning the mean vector and covariance matrix that take into account different aspects of the data.
In this section, we consider an arbitrary transform of a uniform distribution, increasing the expressivity of the attainable sampling distribution.

A well-known result from probability theory states that, given the cumulative distribution function $\fnF_{X}$ of a random variable $\vcx$, then the random variable $\vcu = \fnF_{X}(\vcx)$ is uniformly distributed \cite{Durrett2010-Probability}.
Therefore, defining $\fnF_{X}^{-1}$ as the generalized inverse of $\fnF_{X}$, we can obtain samples from any random variable $\vcx$ by taking samples from a uniform $\vcu$ and transforming those samples through $\fnF_{X}^{-1}$.
Computing $\fnF_{X}^{-1}$, however, is typically intractable.

We propose to learn a transform $\fnPsi_{t}: \fdR^{N} \to \fdR^{N}$ that maps a uniform random variable $\vcu$ into the sample $\vcx_{t}^{(k)}$ for each time $t$.
To do this, we parametrize $\fnPsi_{t}$ by means of a neural network.
In order to include the information from the past state and the current measurement, we consider the following architecture
\begin{equation} \label{eq:arbitraryTransform}
    \vcx_{t}^{(k)} = \fnPsi_{t} (\vcu;\vcx_{t-1}^{(k)}, \vcy_{t}) = \NN_{t}^{\fnPsi} \big( \fnrho \big( \mtA_{t}^{\fnPsi} \vcu+ \mtB_{t}^{\fnPsi} \vcx_{t}^{(k)} + \mtC_{t}^{\fnPsi} \vcy_{t}\big) \big)
\end{equation}
where $\vcu \distro \unifDistro([0,1]^{N})$, $\mtA_{t}^{\fnPsi} \in \fdR^{N \times N}$, $\mtB_{t}^{\fnPsi} \in \fdR^{N \times N}$ and $\mtC_{t}^{\fnPsi} \in \fdR^{N \times M}$.

Each sampled state $\vcx_{t}^{(k)}$ in the trajectory [cf. \eqref{eq:arbitraryTransform}] is distributed according to some probability distribution $\vcx_{t}^{(k)} \distro \fnpi_{t}^{\fnPsi}(\vcx_{t}^{(k)}|\vcx_{t-1}^{(k-1)}, \vcy_{t})$.
A means to evaluate values of $\fnpi_{t}^{\fnPsi}$ is required in order to be able to compute the weight associated to each trajectory [cf. \eqref{eq:unnormalizedWeights}].
Observing that $\fnPsi_{t}$ is continuous, and since $\vcu$ is a continuous random variable, then $\vcx_{t}^{(k)}$ is continuous as well, and thus it can be characterized by a continuous probability density function (pdf).
Provided that $\fnPsi_{t}$ is invertible, this pdf can be computed as follows
\begin{equation}
    \fnpi_{t}^{\fnPsi}(\vcx_{t}^{(k)}|\vcx_{t-1}^{(k)}, \vcy_{t}) = \big| \det\big(\mtJ_{\fnPsi_{t}^{-1}}(\vcx_{t}^{(k-1)})\big) \big|
\end{equation}
since $\fnf_{X}(\fnPsi_{t}^{-1}(\vcx_{t}^{(k)})) = 1$, and where $\mtJ_{\fnPsi_{t}^{-1}} \in \fdR^{N\times N}$ is the Jacobian of $\fnPsi_{t}^{-1}$.
This Jacobian can be computed by noting that
\begin{equation}
    \fnPsi^{-1}(\vcx_{t}^{(k)}) = (\mtA_{t}^{\fnPsi})^{-1}\big( \vcz_{t,0}^{\fnPsi} - \mtB_{t}^{\fnPsi} \vcx_{t-1}^{(k)} - \mtC_{t}^{\fnPsi} \vcy_{t} \big)
\end{equation}
with
\begin{equation}
    \vcz_{t,\ell-1}^{\fnPsi} = (\mtW_{t,\ell}^{\fnPsi})^{-1} \fnrho^{-1}(\vcz_{t,\ell}^{\fnPsi} - \vcb_{t,\ell}^{\fnPsi})
\end{equation}
for $\ell=1,\ldots,L$ the number of layers in the neural network $\NN_{t}^{\fnPsi}$, and where $(\mtW_{t,\ell}^{\fnPsi}, \vcb_{t,\ell}^{\fnPsi})$ are the parameters of the affine transform of layer $\ell$.
Note that $\fnz_{t,L}^{\fnPsi} = \vcx_{t}^{(k)}$.
In other words, $\fnPsi^{-1}$ has a structure analog to that of a neural network, and thus its Jacobian can be computed by an algorithm analog to backpropagation \cite{Rumelhart1986-BackProp}.

While we assumed that $\fnPsi$ is invertible from the moment we decided to use it to learn $\fnF_{X}^{-1}$, neural networks are not necessarily invertible, and thus we need to guarantee this in the design.
The simplest way to do so is to choose square matrices $\mtW_{\ell}^{\Psi}$ for all $\ell$, and thus $\fnF_{\ell} = N$ for all $\ell$.
This curtails the ability of the designer to control the representation capability of the neural network.
We note that a square matrix does not necessarily guarantees invertibility.
One way to approximately compute the inverse, then, would be to add a small identity matrix, which is a common practice in ridge regression estimators \cite{Kay1993-FundamentalsEstimation}.
Also, the activation function has to be invertible, which is the case for the hyperbolic tangent, but not for the ReLU.
More specific solutions can be find in the field of normalizing flows \cite{Kobyzev2020-NormalizingFlows, Papamakarios2021-NormalizingFlows, Eldar2022-UnfoldingNormalizingFlows}.

The set of learnable parameters for each time instant $t$ is given by $\stTheta_{t}^{\fnPsi} = \{ \{\mtW_{t,\ell}^{\fnPsi}, \vcb_{t,\ell}^{\fnPsi}\}_{\ell=1}^{L}, \mtA_{t}^{\fnPsi}, \mtB_{t}^{\fnPsi}, \mtC_{t}^{\fnPsi}\}$.
Note that, in \eqref{eq:arbitraryTransform}, we have chosen to learn a different neural network for each time instant $t$.
This makes the number of learnable parameters a function of $t$, and thus requires more data as the trajectories get longer.
Alternatively, we can fix a single neural network for all $t$, and use the values of $\vcx_{t-1}^{(k)}$ and $\vcy_{t}$ to keep track of the changes in the system across time.


\subsection{Likelihood of the model} \label{subsec:model}

All of the architectures presented so far are determined by a set of parameters.
These parameters are updated iteratively during the training phase by computing gradient descent steps towards optimizing some objective function \cite[Ch. 8]{Goodfellow2016-DeepLearning}.
We train the architectures by choosing to maximize the likelihood of the model.
This results in an unsupervised regime, since only the sequence of measurements $\{\vcy_{t}\}$ is required for training.

The model of the nonlinear dynamical system under study is characterized by the transition distribution $\fnp(\vcx_{t}|\vcx_{t-1})$ and the measurement distribution $\fnp(\vcy_{t}|\vcx_{t})$, both considered known.
Hence, we learn the sampling distributions $\fnpi$ in an unsupervised way by maximizing the likelihood of the model
\begin{equation} \label{eq:likelihood}
    \max_{\stTheta} \sum_{k=1}^{K} \fnp \big( \vcx_{t}^{(k)}[\stTheta] \big| \vcx_{t-1}^{(k)}[\stTheta] \big) \fnp \big(\vcy_{t} \big| \vcx_{t}^{(k)}[\stTheta] \big).
\end{equation}
By the notation $\vcx_{t}^{(k)}[\stTheta]$ we have explicitly indicated that each sample depends on the learnable parameters $\stTheta$ through the learned sampling distribution $\stTheta$.
By updating the parameters $\stTheta$ towards maximizing \eqref{eq:likelihood}, we attempt to generate samples that are reasonable in light of the system dynamics.
For instance, if one sampled value $\vcx_{t}^{(k)}$ would not be reasonable in light of the measurement $\vcy_{t}$, as dictated by the distribution $\fnp(\vcy_{t}|\vcx_{t})$, then the parameters $\stTheta$ are going to be updated, subsequently reducing the probability of sampling such $\vcx_{t}^{(k)}$.
In this way, we can train the architectures using only knowledge of the system dynamics and the measurements $\vcy_{t}$, but with no true knowledge of the states $\vcx_{t}$, which amounts to an unsupervised training regime.

The parameters are typically updated by means of some iterative algorithm based on stochastic gradient ascent \cite{Duchi2011-AdaGrad, Kingma2015-ADAM}.
We note that the objective is not necessarily to optimize the function, but rather to learn to solve the task at hand.
For this reason, usually a fixed number of iterations are done, instead of following some stopping criterion based on the value of the objective function.
Oftentimes, it may be convenient to maximize the log-likelihood, especially when the distributions of the system dynamics belong to the exponential parametric family.
Finally, we remark that since $\vcx_{t}^{(k)}$ is actually a sample drawn from the distribution $\fnpi$ which is the one we want to learn, we use the reparametrization trick to be able to estimate the gradients required for the stochastic gradient ascent algorithm \cite{Kingma2014-VAE}.

\section{Numerical Experiments} \label{sec:sims}

Let us consider a dynamical system given by
\begin{equation} \label{eq:dynamicalSystemExperiment}
\begin{gathered}
    \vcx_{t} = \fnphi(\mtA \vcx_{t-1}) + \vcv_{t}, \\
    \vcy_{t}= \mtC \vcx_{t} +  \vcw_{t},
\end{gathered}
\end{equation}
where $\vcx_{t} \in \fdR^{N}$ is the state, $\vcy_{t} \in \fdR^{M}$ is the measurement, $\mtA \in \fdR^{N \times N}$ is the state transition matrix and $\mtC \in \fdR^{M \times N}$ is the measurement matrix.
{The function $\fnphi: \fdR^{N} \to \fdR^{N}$ may be nonlinear (depending on the simulation scenario).}
The state noise $\vcv_{t} \in \fdR^{N \times N}$ has mean $\xp[\vcv_{t}] = \vcZeros$ and covariance matrix $\xp[\vcv_{t} \vcv_{t}^{\Tr}] = \sigma_{v}^{2} \mtI_{N} \in \fdR^{N \times N}$ for all $t$. Likewise for the measurement noise $\vcw_{t} \in \fdR^{M}$, where $\xp[\vcw_{t}] = \vcZeros$ and $\xp[\vcw_{t} \vcw_{t}^{\Tr}] = \sigma_{w}^{2} \mtI_{N} \in \fdR^{M \times M}$ for all $t$. Noise vectors are independent for any pair $t,t'$ such that $t \neq t'$, and all the random vectors in the sequence $\{\vcv_{t}\}_{t \geq 0}$ are independent from those in the sequence $\{\vcw_{t}\}_{t \geq 0}$.

In this context, we consider to have access to a sequence of $t$ measurements $\{\vcy_{t}\}_{t=0,...,T}$ and we want to estimate the value $\vcx_{t}$ of the state at time $t$, so that the target quantity becomes $\vcz_{t} = \fnf_{t}(\vcx_{0:t}) = \vcx_{t}$. The baseline estimator is then given by $\vctz_{t}$ as in \eqref{eq:conditionalExpectation}. In what follows, we will obtain estimates $\vchz_{t}$ using particle filtering as in \eqref{eq:particleFiltering}, under the consideration of different sampling distributions $\fnpi(\vcx_{t}|\vcx_{0:t-1}^{(k)}, \vcy_{0:t})$ as discussed in Sec.~\ref{sec:unsupervised}.

The baseline sampling distribution is given by the one that minimizes the degeneracy, i.e. $\fnpi(\vcx_{t}|\vcx_{0:t-1}^{(k)}, \vcy_{0:t}) = \fnp(\vcx_{t}|\vcx_{t-1}^{(k)}, \vcy_{t})$. This sampling distribution can be computed in closed form for the dynamical system in \eqref{eq:dynamicalSystemExperiment} for the case in which the noise distributions for $\vcv_{t}$ and $\vcw_{t}$ are Gaussian, see \cite{Doucet2000-ParticleFiltering} for details.

In all the following simulations, we set $M = N-2$ with matrix $\mtA$ being generated as the weighted adjacency matrix of a random geometric graph. The weights are computed using a Gaussian kernel on the distance, i.e. the weight between node $i$ and node $j$ is given by $\exp(-\|\vcr_{i} - \vcr_{j}\|^{2})$ where $\vcr_{i} \in \fdR^{2}$ are the coordinates of node $i$ on the $[0,1]$ plane. A weighted, $3$-nearest neighbor graph is constructed, and the adjacency matrix is normalized so that it has unit spectral norm. The measurement matrix is obtained from $\mtC = \begin{bmatrix} \mtI_{M \times M} & \mtI_{M \times (N-M)} \end{bmatrix}$ where $\mtI_{P \times Q}$ is a $P \times Q$ matrix such that $[\mtI]_{ii} = 1$ and $[\mtI]_{ij} = 0$ for all $i \neq j$, normalized to have unit spectral norm. Such dynamical system is characteristic of diffusion processes in graphs, including rumor spreads, heat diffusion, and graph filtering \cite{Gama2019-LinearControl}.

The initial state $\vcx_{0}$ is drawn from a multivariate Gaussian with mean $\xp[\vcx_{0}] = \vcmu^{0} = \vcOnes_{N}$ and covariance matrix $\xp[(\vcx_{0} - \vcmu^{0})(\vcx_{0} - \vcmu^{0})^{\Tr}] = \mtI_{N}$. In this setting, we define the state noise SNR as $10 \log_{10}(\|\vcmu_{0}\|^{2} / \sigma_{v}^{2})$ and the measurement noise SNR analogously. For the simulations, we consider the SNR to be $5\text{dB}$, fixing the value of both $\sigma_{v}^{2}$ and $\sigma_{w}^{2}$.

We construct particle filter estimates drawing $K$ particles, with resampling whenever $\scK_{t}^{\text{eff}}$ is smaller than $\schK^{\text{thres}} = K/3$. To account for the randomness in the generation of the particle filtering estimate, we repeat the process $100$ times. That is, we sample $K$ particles, construct the estimate, sample another $K$ particles, construct another estimate, and so on for $100$ repetitions. These estimates are then averaged to produce a single particle filtering estimate that accounts for the inherent randomness in the particle filter.

For the learnable sampling distributions leveraging fully-connected neural networks to learn the mean and covariance matrix of a multivariate distribution, we consider $4$ layers, where the input to the first layer is of size $N+M$ and the output of the last layer is of size $N$, as explained in Sec.~\ref{subsec:MLP}. The number of hidden units is set to $256$, $512$ and $1,\!024$ at the outputs of layers $1$, $2$, and $3$, respectively. We use the same hyperparameters for the mean neural network and the covariance neural network (note that the hyperparameters are the same, but the parameters actually learned will be different). For the RNN model (Sec.~\ref{subsec:RNN}), the size of the hidden state is set to $H = 1024$. For the GNN (Sec.~\ref{subsec:GNN}) we consider the graph matrix description to be $\mtS = \mtA$ the state transition matrix, and we use $4$ layers with dimensions $F_{t,1} = 256$, $F_{t,2} = 512$, $F_{t,3} = 1024$ and $F_{t,4} = 1$ for all $t$. The number of neighborhood exchanges is set to $D=3$ for all filters involved.
Finally, for the learnable sampling distribution capable of learning an arbitrary distribution (i.e. not the parameters of a multivariate normal) described in Sec.~\ref{subsec:arbitrary}, we consider $9$ layers (recall that the number of hidden units is fixed by the requirement that the matrices be squared). 
In all learnable architectures, the activation function is set to be the hyperbolic tangent.

\begin{figure*}[t]
    \centering
    \subfloat[$N=10$, $M=8$]{%
        \label{subfig:linearGaussian-N10M8}%
        \includegraphics[width=0.65\columnwidth]{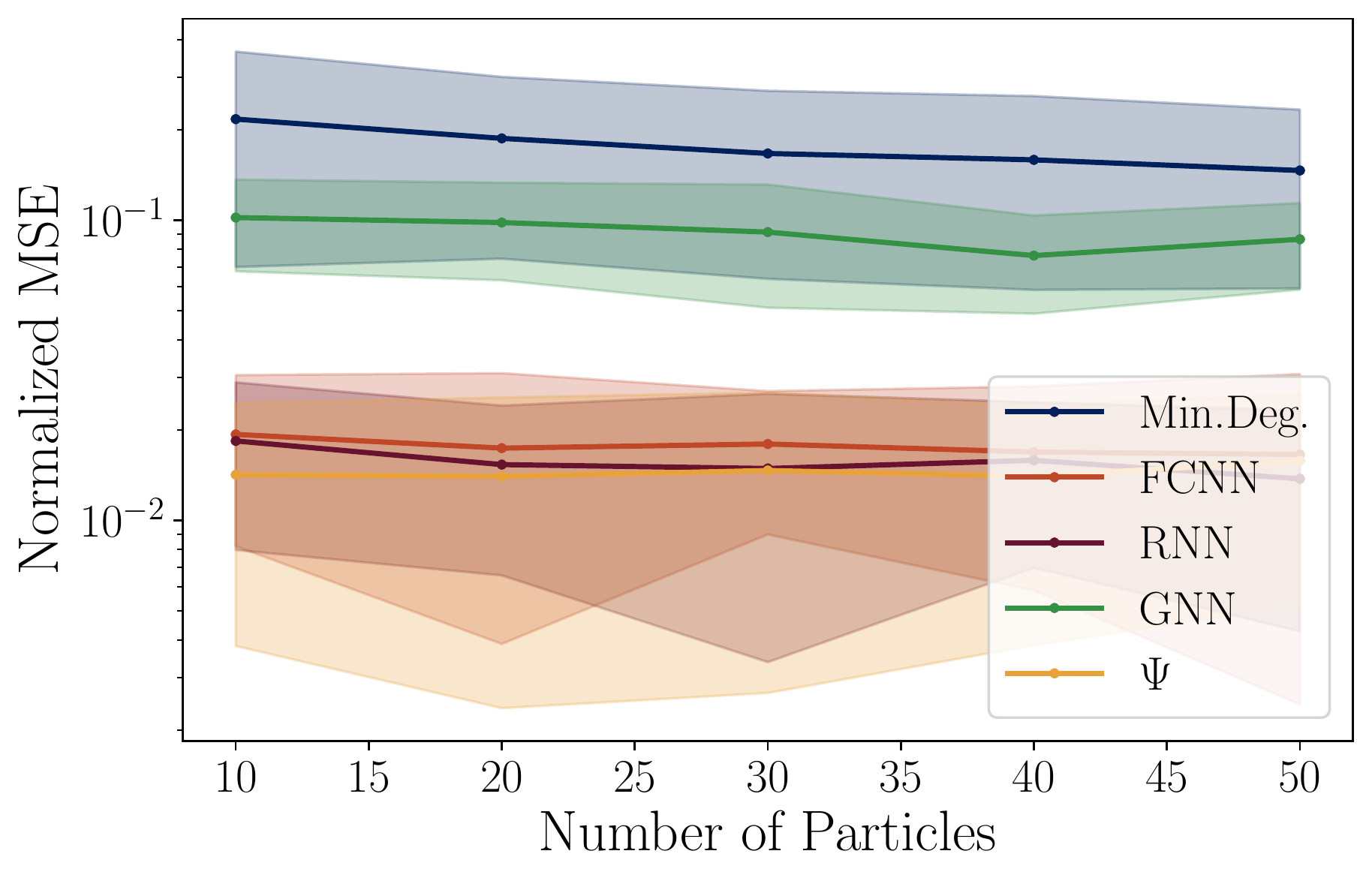}%
    }%
    \hfill
    \subfloat[$N=25$, $M=23$]{%
        \label{subfig:linearGaussian-N25M23}%
        \includegraphics[width=0.65\columnwidth]{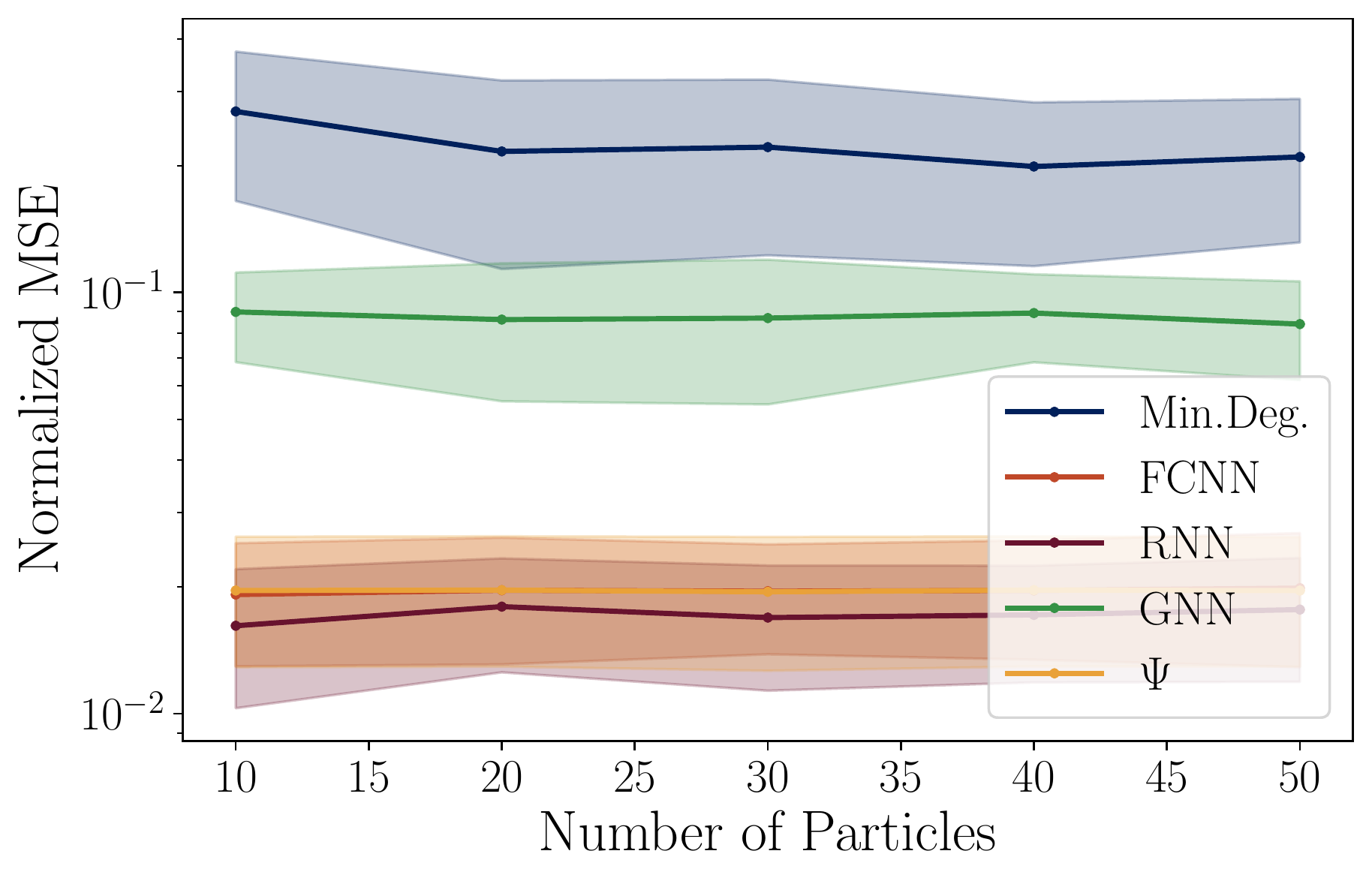}%
    }%
    \hfill
    \subfloat[$N=50$, $M=48$]{%
        \label{subfig:linearGaussian-N50M48}%
        \includegraphics[width=0.65\columnwidth]{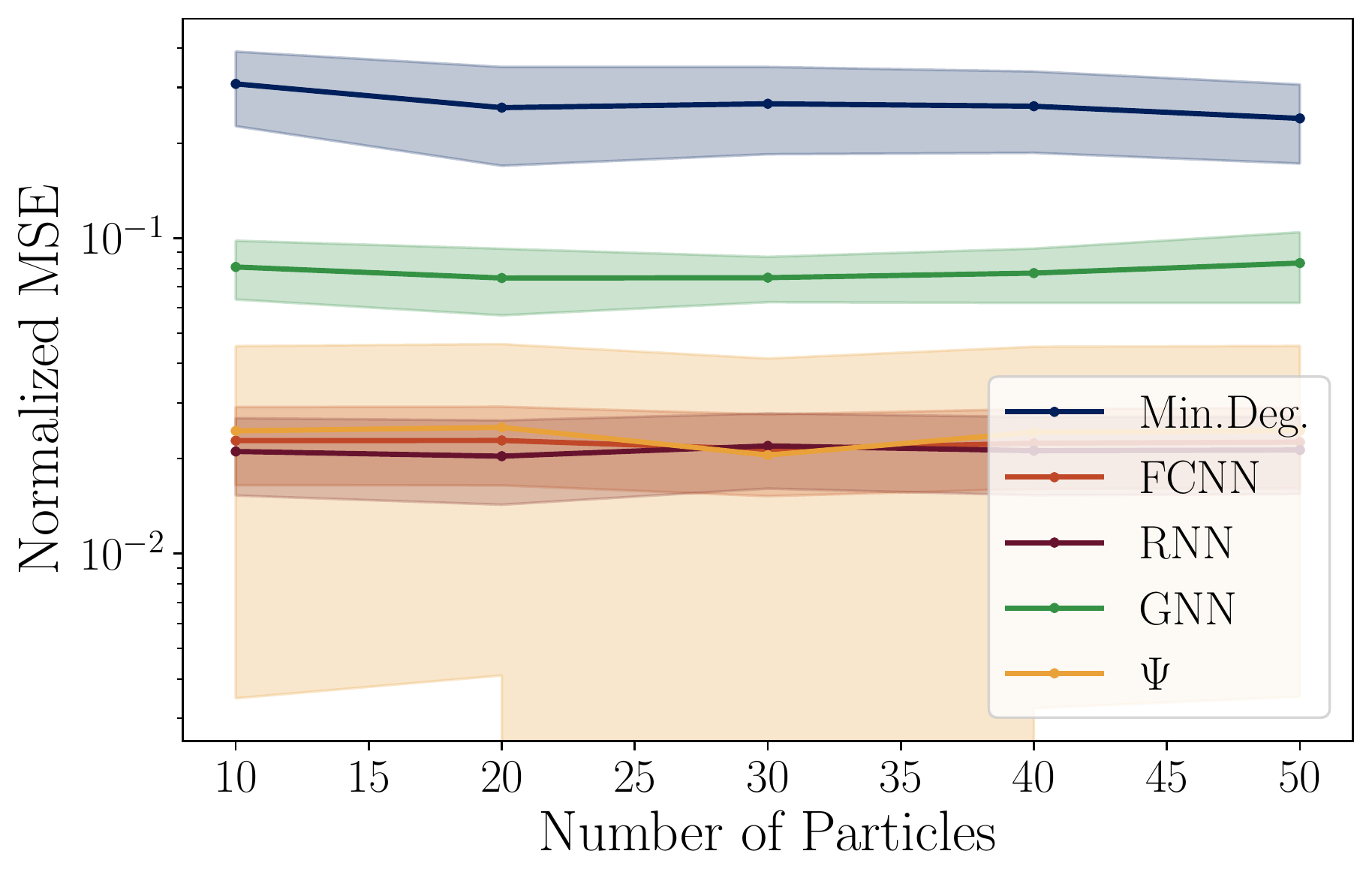}%
    }%
    \caption{Linear Gaussian dynamical system as a function of the number of particles $K$. \protect\subref{subfig:linearGaussian-N10M8}--\protect\subref{subfig:linearGaussian-N50M48} We vary the state dimension $N$ and the measurement dimension $M$. 
    All the sampling distributions that were learned from data significantly outperform the baseline of the designed, minimum-degeneracy sampling distribution.}
    \label{fig:linearGaussian-nParticles}
\end{figure*}

To train these learnable sampling distributions, we have access to a sequence of $t$ measurements $\{\vcy_{t}\}_{t}$ and we find the parameters that maximize the likelihood of the model as described in Sec.~\ref{subsec:model}. To do so, we use an optimization algorithm known as Adam \cite{Kingma2015-ADAM} which is a momentum-based variant of stochastic gradient descent. We use a learning rate of $0.001$, and forgetting factors of $0.9$ and $0.999$. We carry out $200$ training steps, where each step uses all data (also known as $200$ epochs). To account for the randomness of the system dynamics (in generating the state transition matrices $\mtA$) and of the generated measurements $\{\vcy_{t}\}$ we repeat the whole learning and testing process $20$ times. We report the median performance of each particle filter estimate, together with the corresponding standard deviation.

\begin{figure}[t]
    \centering
    \includegraphics[width=0.75\columnwidth]{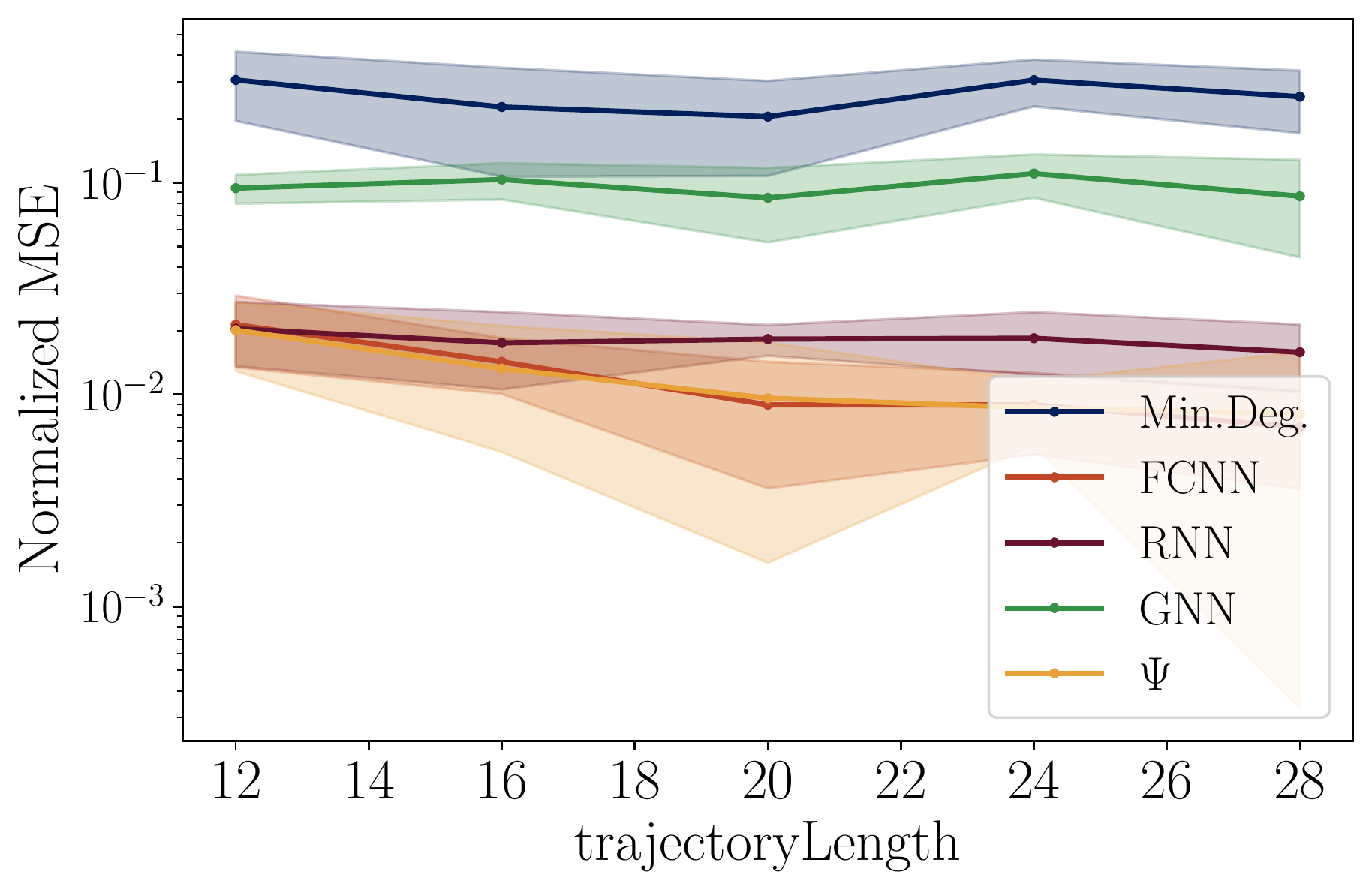}%
    \caption{Linear Gaussian dynamical system as a function of the trajectory length $T$.}
    \label{fig:linearGaussian-trajectoryLength}
\end{figure}

In the following simulations, we explore how the performance of particle filters with learnable sampling distributions changes as a function of the state dimension $N$ (recall that the measurement dimension is set to $M = N-2$), the trajectory length $T$, the number of particles $K$, the noise distribution, and the nonlinear function $\fnphi$.


\subsection{Linear Gaussian dynamical system} \label{subsec:linear}

First, we consider a linear dynamical system, where $\fnphi$ is the identity function
\begin{equation} \label{eq:LinearSystemExperiment}
    \begin{gathered}
        \vcx_{t} = \mtA \vcx_{t-1} + \vcv_{t}, \\
        \vcy_{t}= \mtC \vcx_{t} +  \vcw_{t},
    \end{gathered}
\end{equation}
and we also consider both the state noise and the measurement noise to be multivariate Gaussian.
In such a scenario, we can compute the ground truth estimator in closed form $\xp[\vcx_{t}|\vcy_{0:t}]$ since $\vcx_{t}|\vcy_{0:t}$ is also a multivariate Gaussian random variable.

\begin{figure}[t]
    \centering
    \includegraphics[width=0.75\columnwidth]{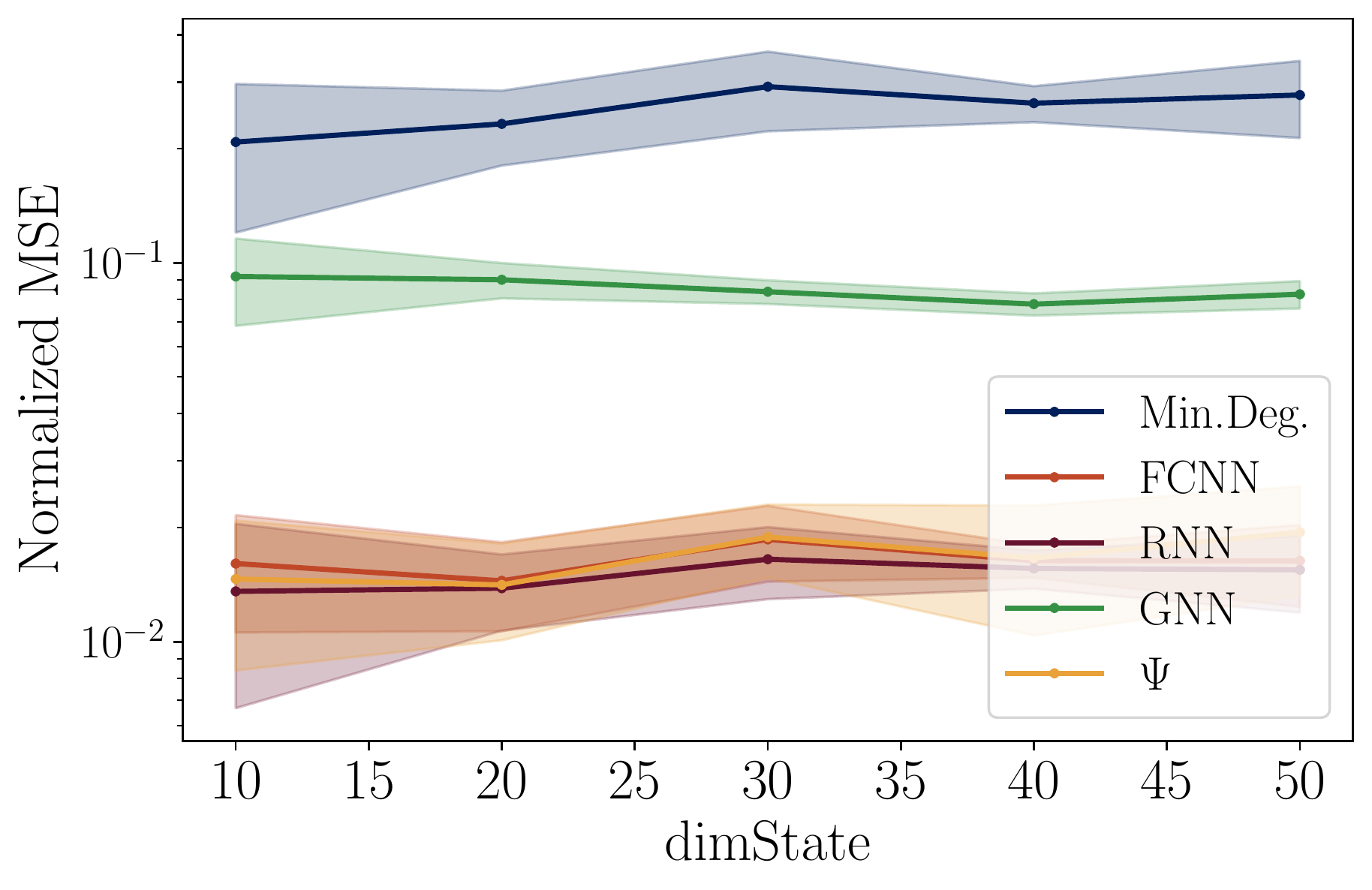}%
    \caption{Linear Gaussian dynamical system as a function of the state dimension $N$.}
    \label{fig:linearGaussian-dimState}
\end{figure}

We consider three different values of state dimension, i.e. $N = 10$, $N=25$ and $N=50$. The resulting, normalized mean squared error between the particle filter estimators and the ground truth is plotted in Fig.~\ref{fig:linearGaussian-nParticles} as a function of the number of particles in the set $K \in \{10, 20, 30, 40, 50\}$. In all cases, we set the trajectory length to be $T=12$.

For all the state dimensions, we observe that the learned sampling distributions considerably outperform the designed, minimum degeneracy sampling distribution. 
We note that there is no significant difference between using a fully connected neural network (Sec.~\ref{subsec:MLP}), a recurrent neural network (Sec.~\ref{subsec:RNN}) or an arbitrary learned linear transform. 
Interestingly enough, the arbitrary linear transform $\fnPsi$ exhibits comparable performance to the FCNN and the RNN, even though it relies on significantly less number of parameters. One difference is that the arbitrary linear transform $\fnPsi$ exhibits higher variance, especially for the large case of $N=50$.
The sampling distribution based on the GRNN still performs better than the minimum-degeneracy designed sampling distribution, but considerably worse than the other learnable distributions.
We believe this may be caused by overfitting due to already incorporating information on the underlying graph $\mtA$ in the form of the matrix $\mtS$ used in the graph filtering layers.

Next, we fix the state dimension to be $N=10$, the number of measurements to be $M=8$, and the number of particles to be $K=10$, and we simulate the performance of the particle filters with learnable distributions as a function of $T \in \{12, 16, 20, 24, 28\}$.

\begin{figure*}[t]
    \centering
    \subfloat[$N=10$, $M=8$]{%
        \label{subfig:nonlinearGaussian-N10M8}%
        \includegraphics[width=0.65\columnwidth]{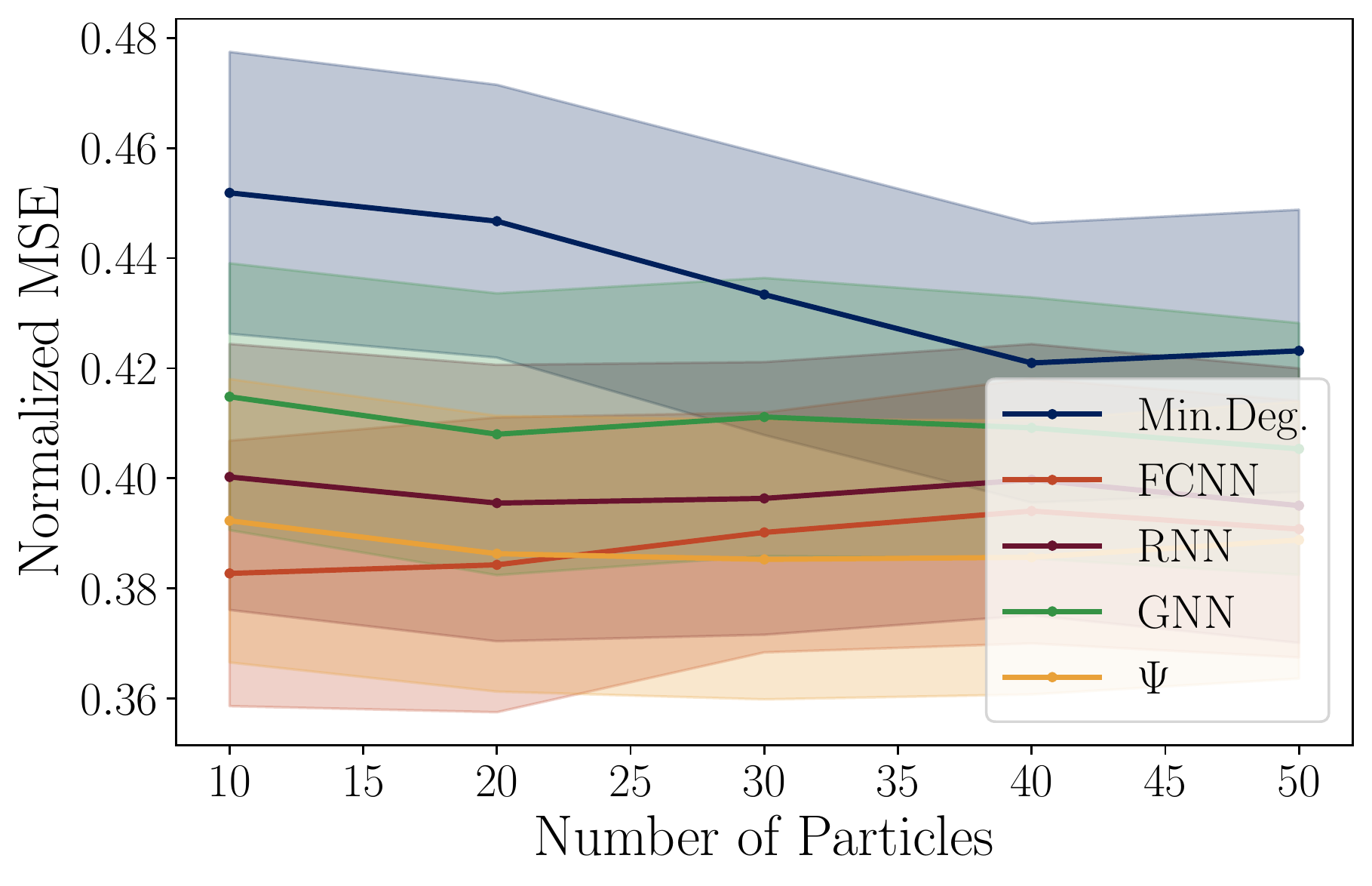}%
    }%
    \hfill
    \subfloat[$N=25$, $M=23$]{%
        \label{subfig:nonlinearGaussian-N25M23}%
        \includegraphics[width=0.65\columnwidth]{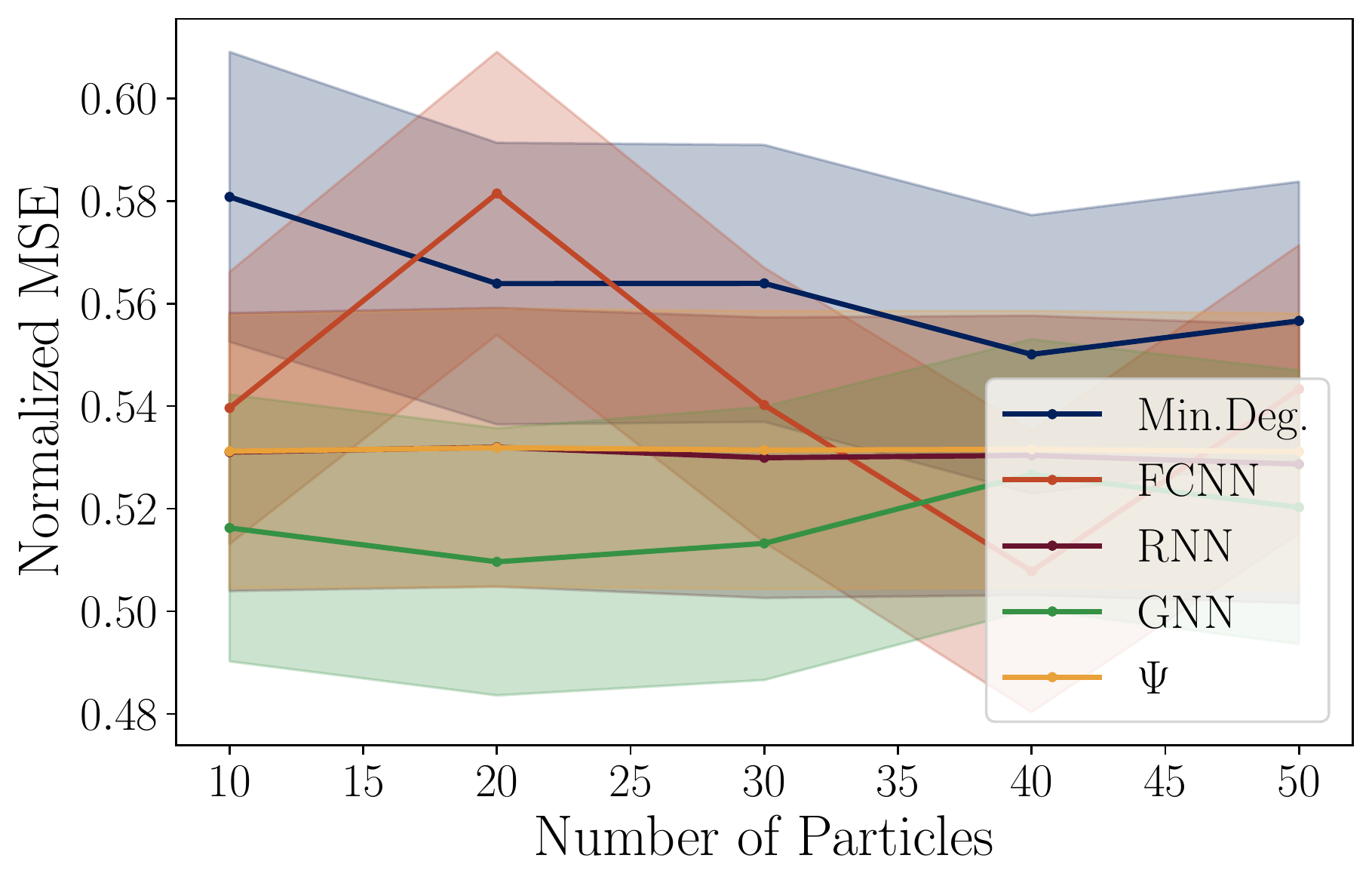}%
    }%
    \hfill
    \subfloat[$N=50$, $M=48$]{%
        \label{subfig:nonlinearGaussian-N50M48}%
        \includegraphics[width=0.65\columnwidth]{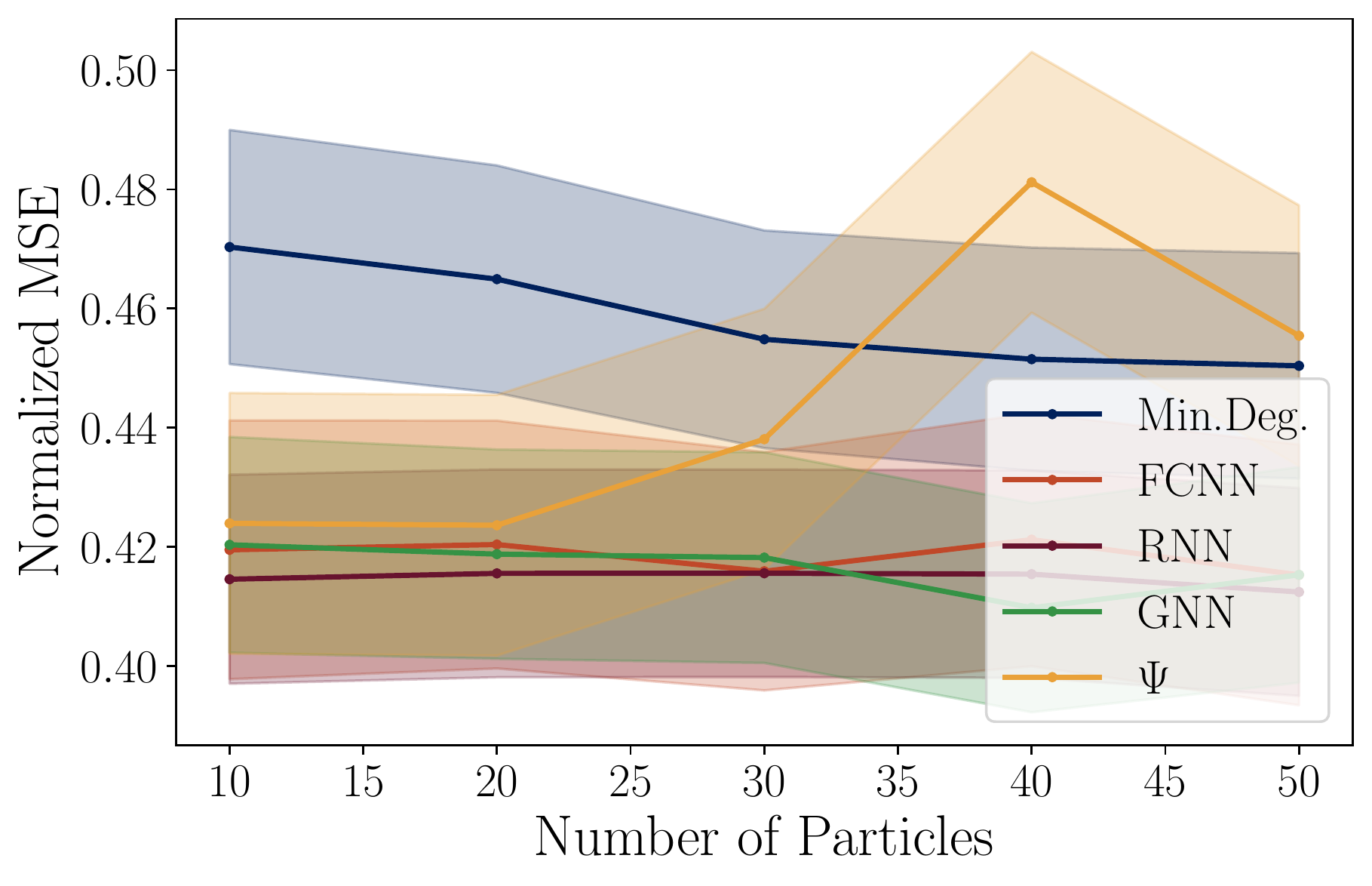}%
    }%
    \caption{Nonlinear Gaussian dynamical system as a function of the number of particles $K$. \protect\subref{subfig:nonlinearGaussian-N10M8}--\protect\subref{subfig:nonlinearGaussian-N50M48} We vary the state dimension $N$ and the measurement dimension $M$. 
    While, generally, the learned distributions outperform the minimum-degeneracy one, the relative gap is now smaller compared to Fig.~\ref{fig:linearGaussian-nParticles}. Additionally, for some distributions (like the FCNN or the arbitrary transform $\fnPsi$) the performance is comparable to the minimum degeneracy (and may be worse) for some values of $K$.}
    \label{fig:nonlinearGaussian}
\end{figure*}

Results are shown in Fig.~\ref{fig:linearGaussian-trajectoryLength}. 
Again, we note that the particle filter with learned sampling distribution outperforms the use of a designed, minimum-degeneracy one, with the GNN performing worse than the other three alternatives.
We note that the performance of the RNN seems independent of the trajectory length, probably due to the fact that this the parameters of the RNN are time-independent by design.
Meanwhile, the performance of the FCNN and the arbitrary linear transform $\fnPsi$ improve with trajectory length, likely because of the increased expressivity that is obtained by learning a different set of parameters for each time instant.
In any case, the performance of these three architectures is comparable.

Finally, we fix the trajectory length to be $T=12$ and the number of particles to be $K=10$ and we simulate as a function of the state dimension for $N \in \{10, 20, 30, 40, 50\}$ (recall that, for each case, $M=N-2$).

Results shown in Fig.~\ref{fig:linearGaussian-dimState} exhibit a similar behavior as in the previous experiments, where all learned distributions perform better than the designed, minimum-degeneracy one, and where the GNN works worse than the rest.
In this case, the performance seems to be independent of the dimension of the state $N$, with all three (FCNN, RNN and arbitrary transform $\fnPsi$) performing similarly.


\subsection{Nonlinear Gaussian dynamical system} \label{subsec:nonlinear}

Next, we consider a nonlinear dynamical system, where $\fnphi$ is set to be the absolute value. The distributions of the state noise and the measurement noise remain Gaussian.
In this scenario, the ground truth $\xp[\vcx_{t}|\vcy_{0:t}]$ can no longer be computed in closed-form. In this case, we use the simulated values of $\vcx_{t}$ as ground truth to measure performance against, i.e. the results shown in the figures report the normalized mean squared error between the different methods and the estimated ground truth value $\vcx_{t}$.

For this case, we fix $N=10$, $M=8$ and $T=12$, and present simulations as a function of the number of particles $K \in \{10, 20, 30, 40, 50\}$. Results are shown in Fig.~\ref{fig:nonlinearGaussian}.

\begin{figure*}[t]
    \centering
    \subfloat[$N=10$, $M=8$]{%
        \label{subfig:linearExponential-N10M8}%
        \includegraphics[width=0.65\columnwidth]{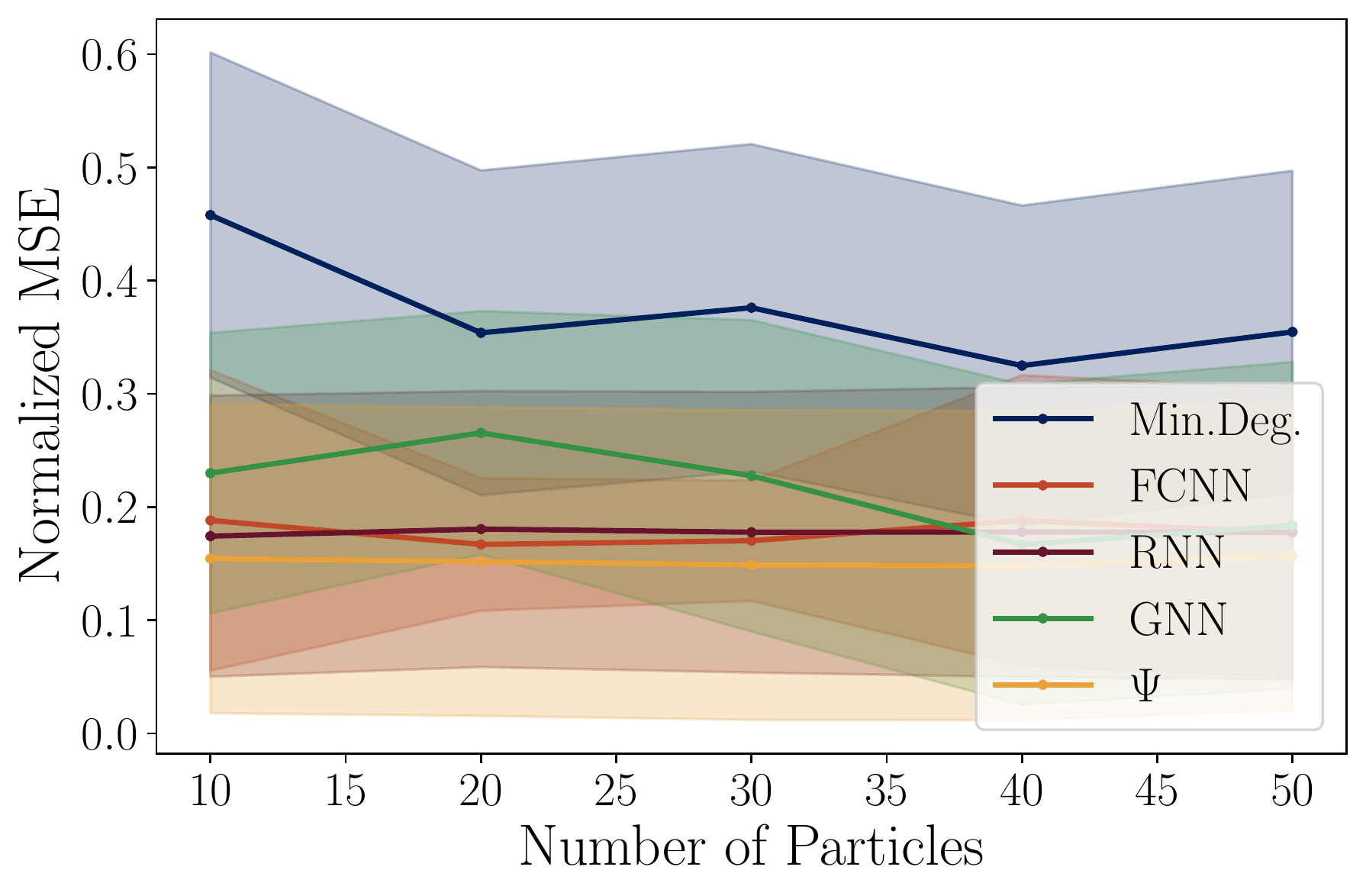}%
    }%
    \hfill
    \subfloat[$N=25$, $M=23$]{%
        \label{subfig:linearExponential-N25M23}%
        \includegraphics[width=0.65\columnwidth]{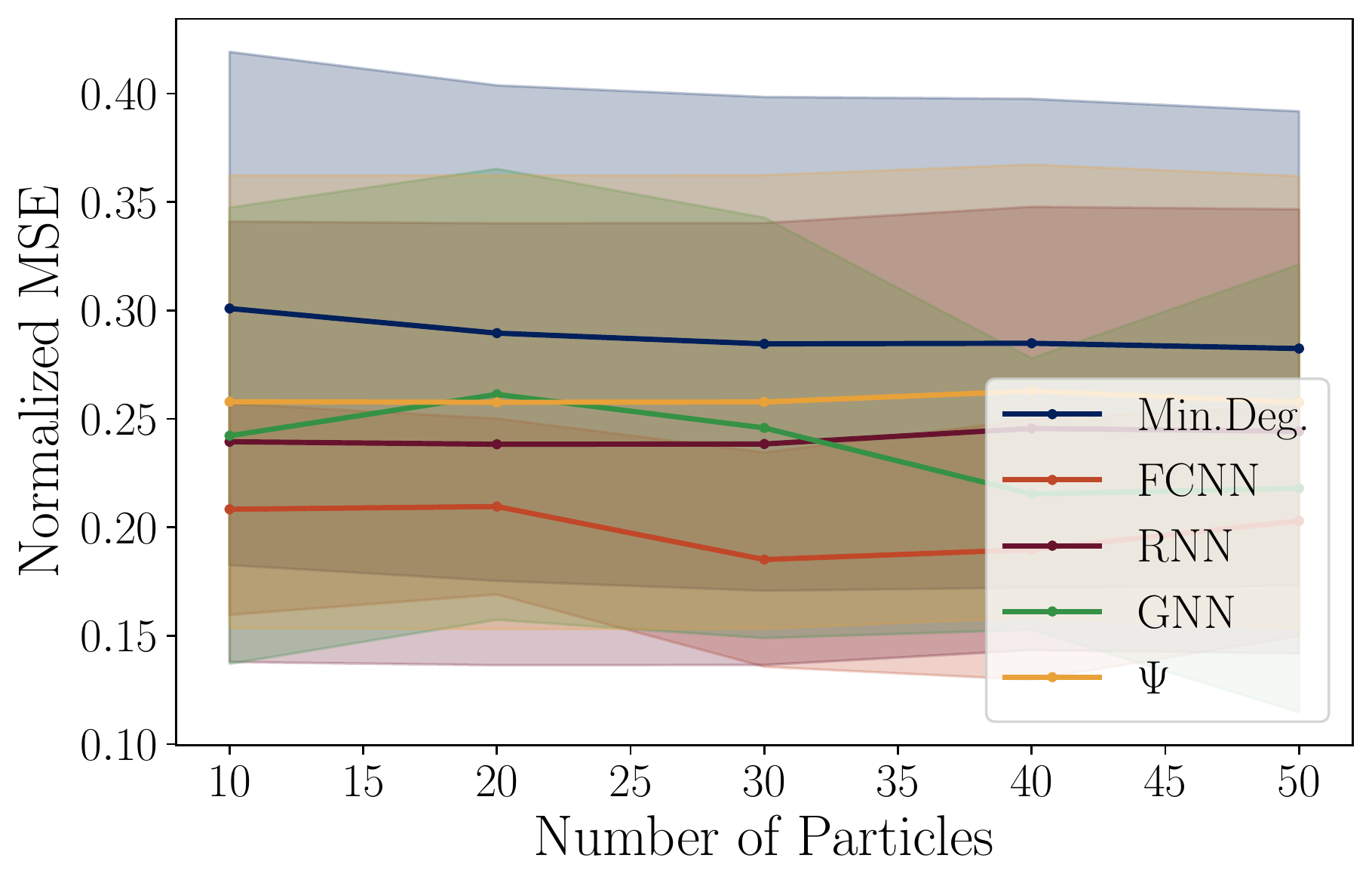}%
    }%
    \hfill
    \subfloat[$N=50$, $M=48$]{%
        \label{subfig:linearExponential-N50M48}%
        \includegraphics[width=0.65\columnwidth]{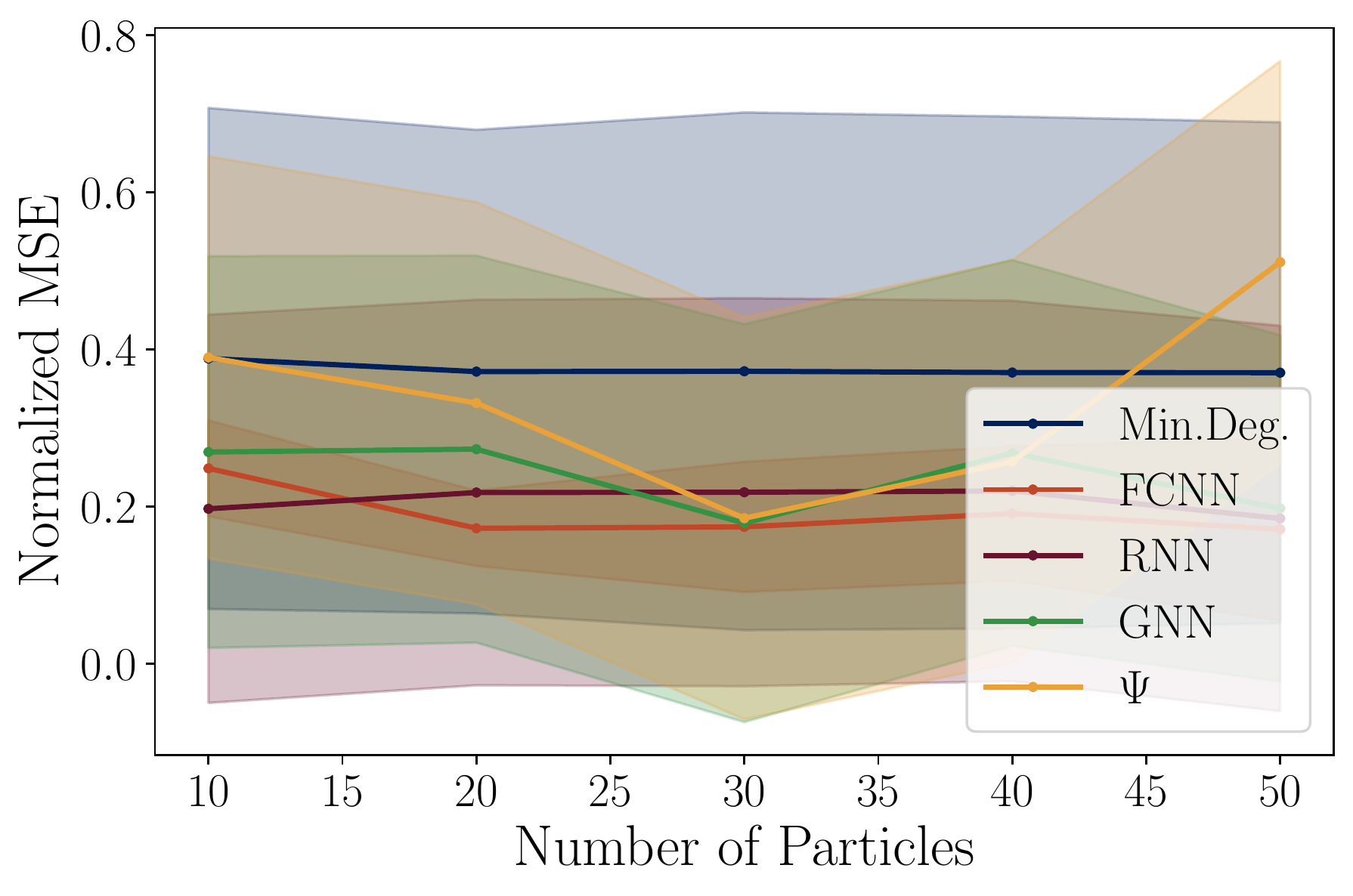}%
    }%
    \caption{Linear non-Gaussian dynamical system as a function of the number of particles $K$. \protect\subref{subfig:linearExponential-N10M8}--\protect\subref{subfig:linearExponential-N50M48} We vary the state dimension $N$ and the measurement dimension $M$. 
    The performance of the arbitrary distribution $\fnPsi$ becomes comparable to the one of the minimum degeneracy as the dimension $N$ of the system grows.}
    \label{fig:linearExponential}
\end{figure*}

The first observation is that, while in general the learned sampling distributions outperform the designed, minimum-degeneracy one, the gap now is significantly smaller.
Furthermore, the behavior of learned distributions such as the FCNN or the arbitrary linear transform $\fnPsi$ is slightly more erratic as a function of the number of particles $K$, sometimes being comparable to the minimum-degeneracy particle filter, sometimes being slightly worse.
This is likely due to the added complexity of the model, and the inability of these generic architectures of adequately capturing the information without proper regularization.
As a matter of fact, the impact of regularization (i.e. choosing operations that reflect certain structure, being either time, like RNNs, or graphs, like GNNs) is evident in the fact that the performance of GNNs is now significantly better (in relative terms) and that both the GNN and the RNN offer a stable behavior as a function of $K$.


\subsection{Linear non-Gaussian dynamical system} \label{subsec:nongaussian}

\begin{figure*}[t]
    \centering
    \subfloat[$N=10$, $M=8$]{%
        \label{subfig:linearUniform-N10M8}%
        \includegraphics[width=0.65\columnwidth]{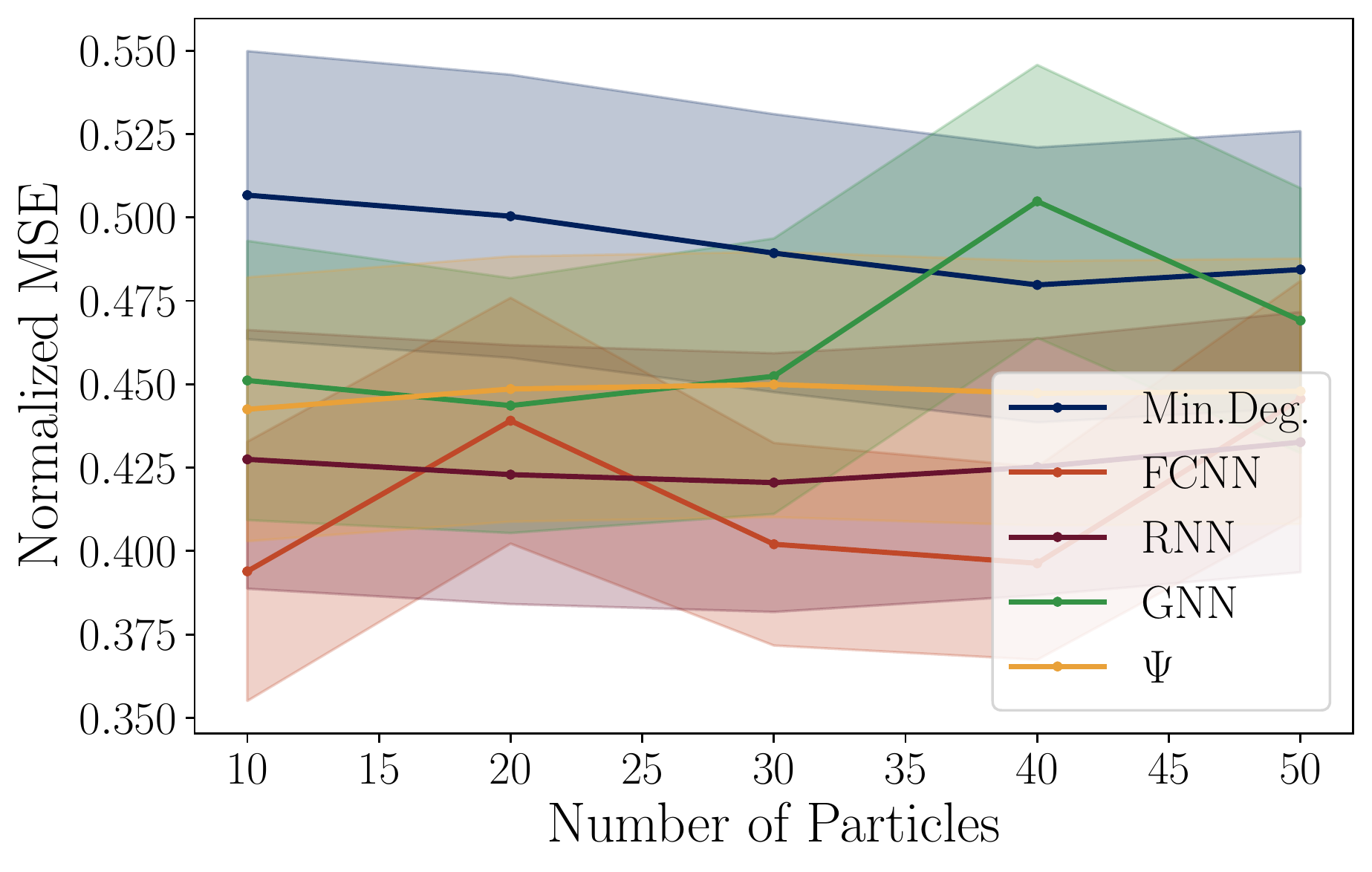}%
    }%
    \hfill
    \subfloat[$N=25$, $M=23$]{%
        \label{subfig:linearUniform-N25M23}%
        \includegraphics[width=0.65\columnwidth]{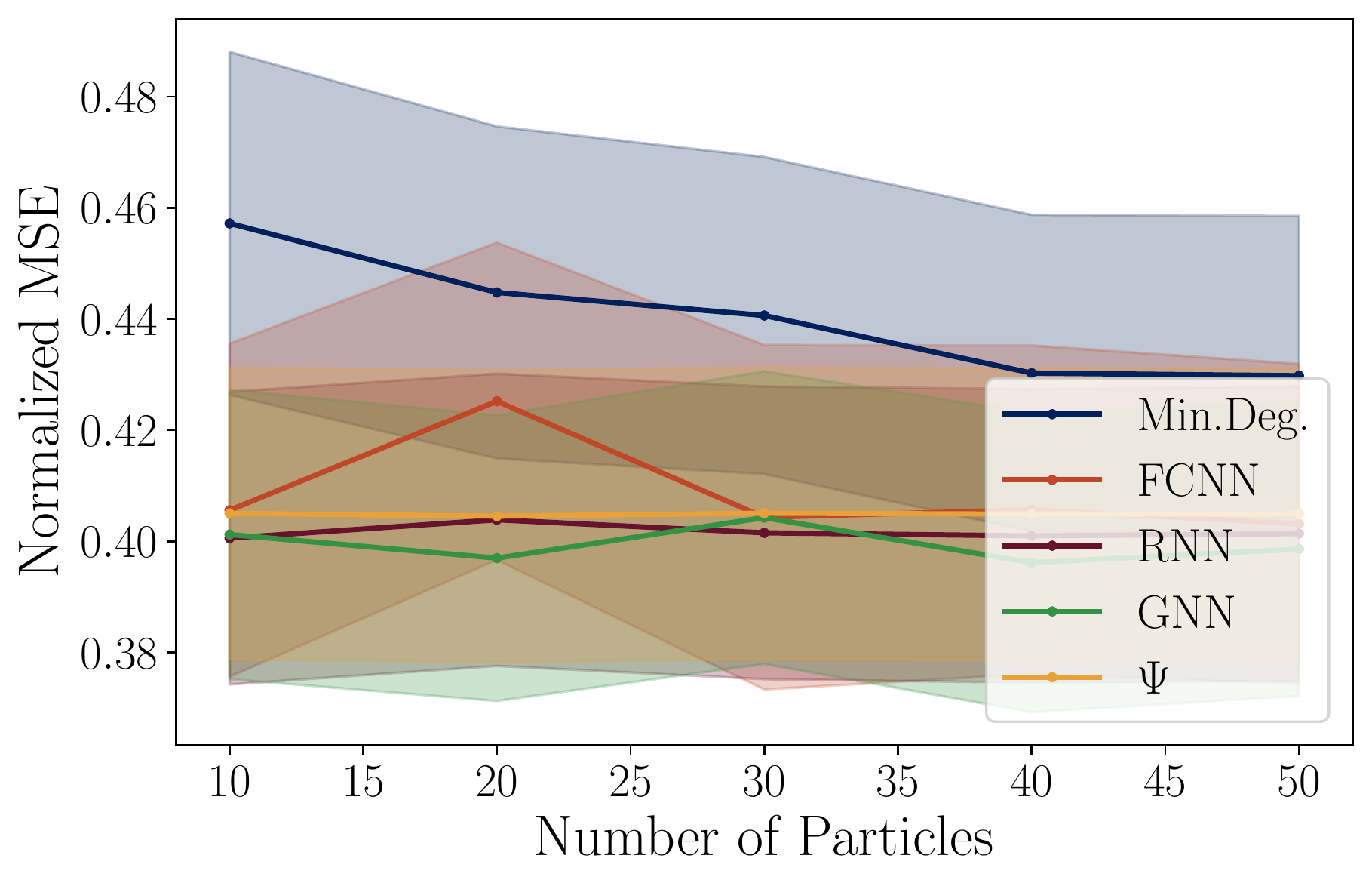}%
    }%
    \hfill
    \subfloat[$N=50$, $M=48$]{%
        \label{subfig:linearUniform-N50M48}%
        \includegraphics[width=0.65\columnwidth]{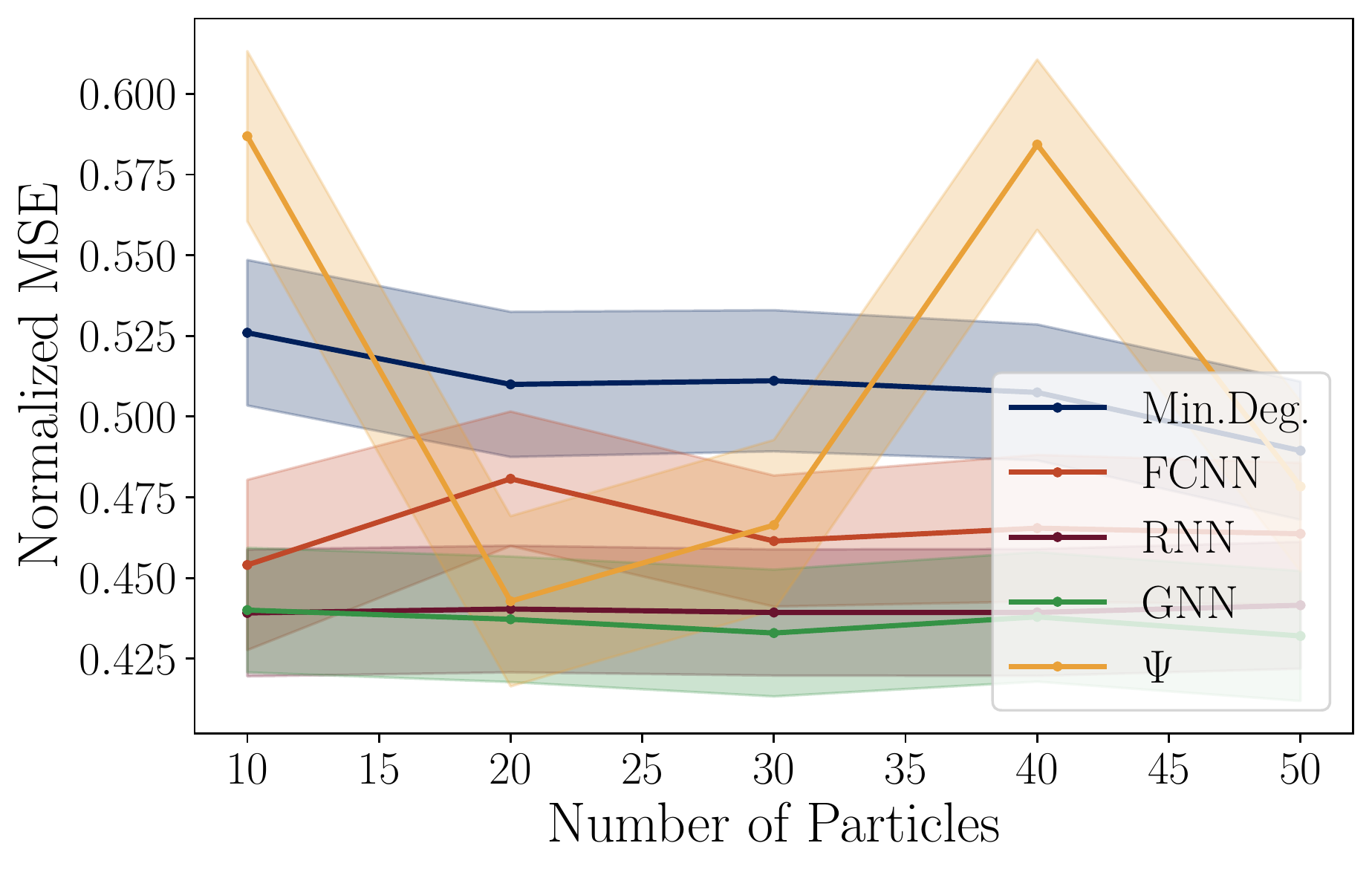}%
    }%
    \caption{Linear-Exponential Dynamical System as a function of the number of particles $K$. \protect\subref{subfig:linearUniform-N10M8} System with state dimenion $N=10$ and measurement dimension $M=8$. \protect\subref{subfig:linearUniform-N25M23} System with state dimension $N=25$ and measurement dimension $M=23$. \protect\subref{subfig:linearUniform-N50M48} System with state dimension $N=50$ and measurement dimension $M=48$. While, generally, the learned distributions outperform the minimum-degeneracy one, the gap is now smaller. Additionally, the performance of the arbitrary distribution $\fnPsi$ becomes comparable to the one of the minimum degeneracy as the dimension $N$ of the system grows.}
    \label{fig:linearUniform}
\end{figure*}

For the third simulation we consider a linear dynamical system \eqref{eq:LinearSystemExperiment}, but with non-Gaussian state and measurement noise. 
Again, we fix $N=10$, $M=8$ and $T=12$, and present simulations as a function of the number of particles  $K \in \{10, 20, 30, 40, 50\}$. 
Results are shown in Fig.~\ref{fig:linearExponential} for exponential noise and in Fig.~\ref{fig:linearUniform} for uniform noise. In both cases, the mean and covariance matrix are still given by the values described at the beginning of the section, but the noise samples are sampled from the corresponding distributions.

As in Sec.~\ref{subsec:nonlinear}, the learned sampling distributions outperform the designed, minimum-degeneracy one, by a gap smaller than the one observed in the linear Gaussian case (Sec.~\ref{subsec:linear}).
We note that the FCNN now exhibits a much more stable behavior than in the nonlinear Gaussian case (Sec.~\ref{subsec:nonlinear}), suggesting that the linearity of the system is more important in the performance of the FCNN than the distribution of the noise.
Here, the performance of the arbitrary linear transform $\fnPsi$ is considerably more erratic, making it somewhat unreliable.
Regularization techniques may come in handy to avoid this erratic behavior.
Finally, we note that the RNN is the more consistent performer with the most stable behavior.

\subsection{Nonlinear non-Gaussian dynamical system} \label{subsec:sir}

We consider a mathematical model that is often employed to describe the dynamics of an epidemic: the SIR model.
It is described by a system of ordinary differential equations~\cite{Beckley2013ModelingEW} that can be discretized via the Euler method for our purposes. 
By doing so with a time step of size $\Delta$, the discrete system is given by
\begin{equation} \label{eq:sir}
    \begin{split}
        S_{t+1} &= - \beta S_t I_t\Delta  + S_t \\
        I_{t+1} &= \left(\beta S_t I_t - \gamma I_t\right) \Delta  + I_t \\
        R_{t+1} &= \gamma I_t  \Delta  + R_t
    \end{split}
\end{equation}
where $S_t$ is the number of susceptible people at time $t$, $I_t$ is the number of infected people at time $t$, and $R_t$ represents the number of people that have been removed from the system (either by death or recovery) by time $t$. 
The parameters of the model $\beta$, $\gamma$, and $\Delta$ are assumed known.
The state of the system is given by $\vcx_t = [S_t, I_t, R_t]^{\Tr}$. 
Notice that the system described in~\eqref{eq:sir} can be thought of as the nonlinear function $\fnphi$ within the framework defined in~\eqref{eq:dynamicalSystemExperiment}.

\begin{figure*}[t]
    \centering
    \subfloat[{$t \in [0, 200]$}]{%
        \label{subfig:sir_0_200}%
        \includegraphics[width=0.65\columnwidth]{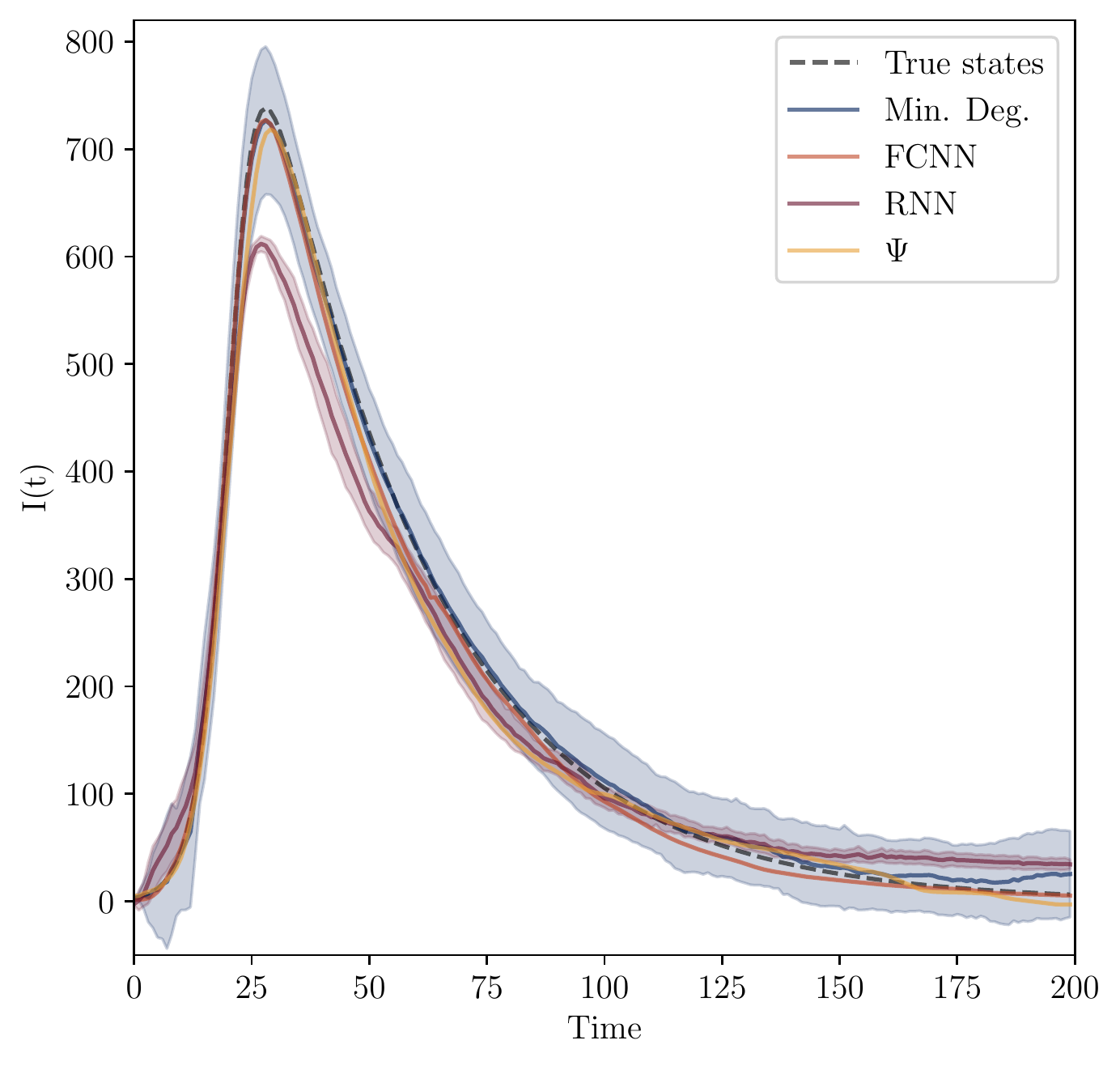}%
    }%
    \hfill
    \subfloat[{$t \in [20, 50]$}]{%
        \label{subfig:sir_20_50}%
        \includegraphics[width=0.65\columnwidth]{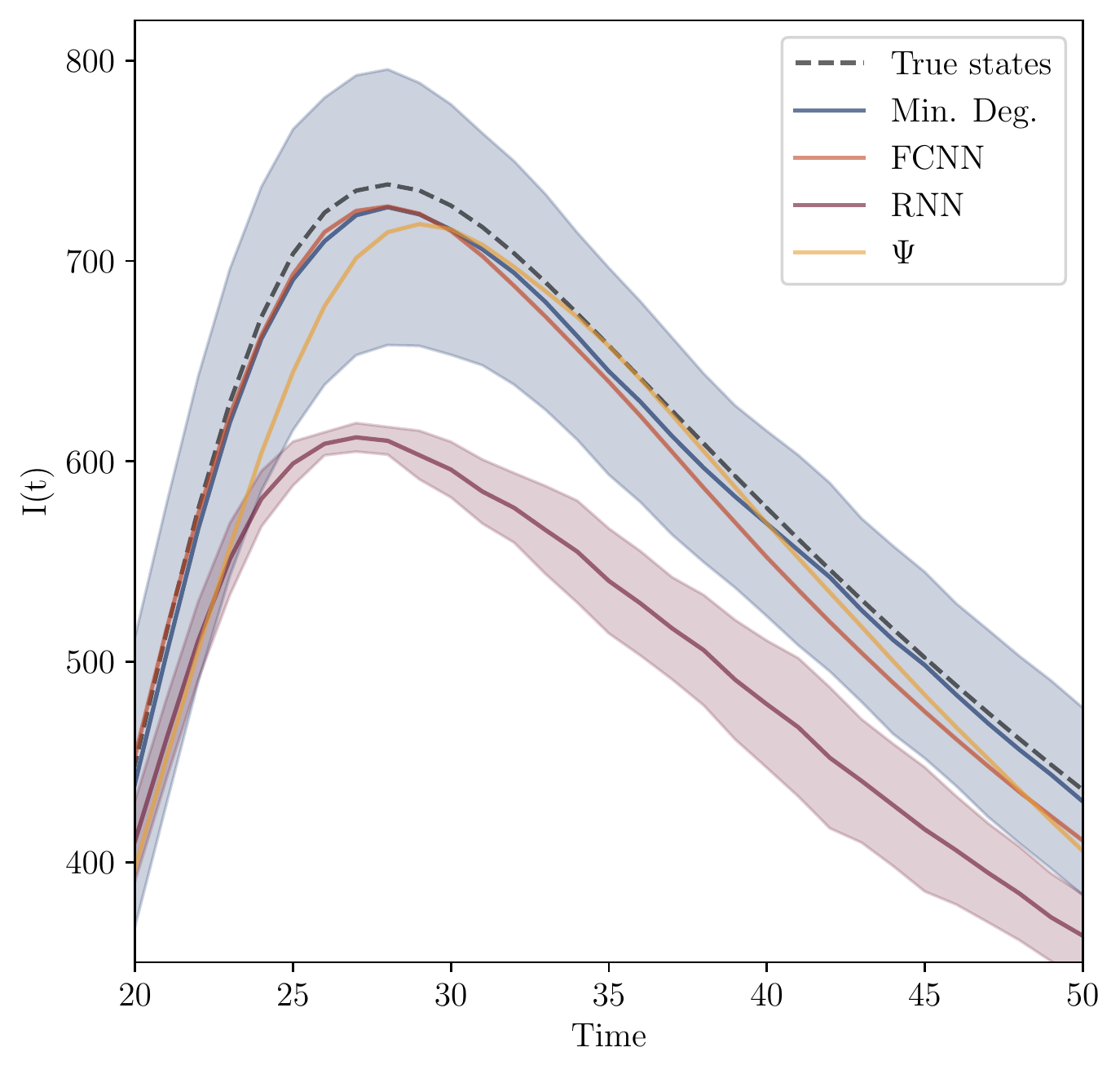}%
    }%
    \hfill
    \subfloat[{$t \in [100, 200]$}]{%
        \label{subfig:sir_100_200}%
        \includegraphics[width=0.65\columnwidth]{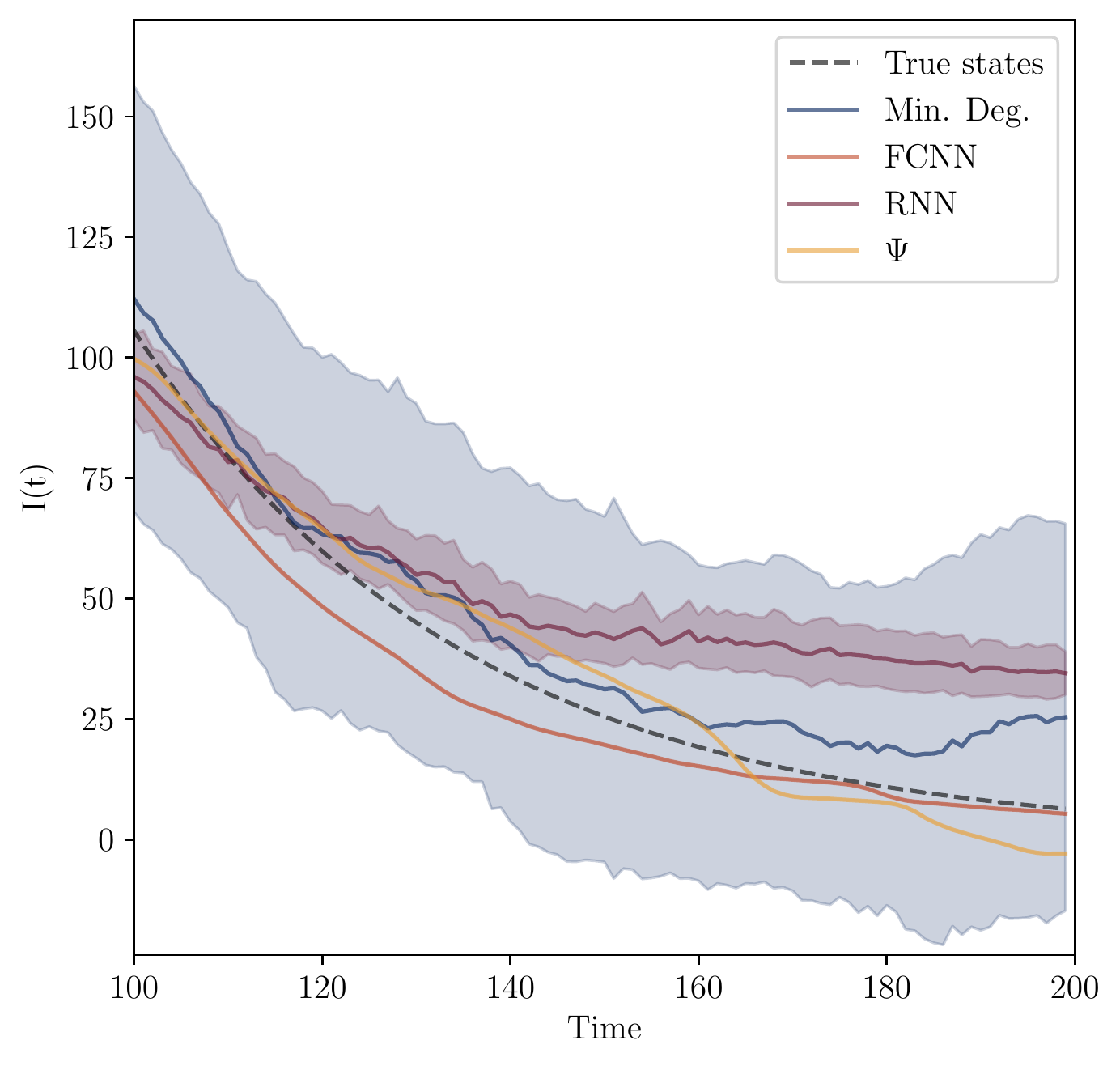}%
    }%
    \caption{The unobserved state $I(t)$ and the estimations using the proposed methods. \protect\subref{subfig:sir_0_200} The state $I(t)$ during all the trajectory.
    \protect\subref{subfig:sir_20_50} The state $I(t)$ zoomed to show the time window where $t \in [20, 50]$.
    \protect\subref{subfig:sir_100_200} The state $I(t)$ zoomed to show the time window where $t \in [100, 200]$. The performance achieved by the FCNN-based method and the arbitrary distribution $\fnPsi$ is higher than that of the minimum degeneracy.}
    \label{fig:sir_comparison}
\end{figure*}

For this experiment, we fix the parameters to be $\beta=5\times10^{-4}$ and $\gamma=0.04$. 
We set a trajectory of $200$ samples, using a time step $\Delta =0.7$. 
We measure $M=2$ states, with $I_t$ being the state that we do not measure. 
More precisely, $\mtC$ in~\eqref{eq:dynamicalSystemExperiment} consists of the first and third row of an identity matrix of size three.
The initial conditions are independently distributed and follow an exponential distribution, with means $\xp[S_0]=997$, $\xp[I_0]=3$ and $\xp[R_0]=0$, and with variances $\mathrm{Var}(S_0)=\mathrm{Var}(I_0)=\mathrm{Var}(R_0)=500$.
The measurement and state noises are also exponentially distributed, with variances $\mathrm{Var}(\vcw_{t})=2500$ and $\mathrm{Var}(\vcv_{t}) = 200$, respectively.
The number of particles is fixed at $K=300$. 
The architecture of the neural networks is changed as well, using $4$ hidden layers in the case of the FCNN (with $512$, $256$, $128$ and $32$ hidden units each, respectively), setting $H=2048$ in the case of the RNN, and using 10 layers for the arbitrary distribution $\fnPsi$. 
The GNN-based method is not used in this experiment, as there is no graphical structure in the system dynamics that could be exploited. 
The estimation of the unobserved state using the different methods is shown in Fig.~\ref{fig:sir_comparison}. 
It is also of interest to evaluate how the learned sampling distributions perform in terms of filtering out the noise from the measured states.
The $\mathrm{MSE}$ for each method can be seen in Table~\ref{tab:sir_comparison}.

The FCNN-based method results to be the one that achieves the best performance, not just at estimating the unknown state but also at filtering the noise from the measured ones.
This method offers the lowest variance and also the lowest bias.
The arbitrary distribution $\fnPsi$ outperforms the minimum-degeneracy one too, although the bias is somewhat larger than the one attained using the FCNN.
Nonetheless, the variance of the estimator is really low as well.
The RNN-based method fails to provide a sampling distribution whose performance is comparable to the other two proposed methods.
Not only is the variance much higher (although still lower than in the minimum-degeneracy case), but the bias is not desirable either.
This can be clearly observed in Fig.~\ref{subfig:sir_20_50}, where the peak of the signal $I(t)$ is correctly captured by the other three methods but not by the RNN-based one.

\begin{table}\centering
\caption{$\mathrm{MSE}$ between the true states and the estimates. 
}
\begin{tabular}{cccc}
        & $S(t)$ & $I(t)$ & $R(t)$ \\
\hline
Min. Deg.  & $419.1$ & $1943.3$ & $538.0$ \\
FCNN & $64.1$ & $136.1$ & $190.8$ \\
RNN & $453.5$ & $2184.6$ & $426.6$\\
$\fnPsi$ & $339.8$ & $453.0$ & $154.6$\\
\hline
\end{tabular}
\label{tab:sir_comparison}
\end{table}

\section{Conclusions} \label{sec:conclusions}

We proposed four different methods for learning the sampling distribution of a particle filter.
The first three are parametric methods, which sample from a multivariate Gaussian distribution with mean and covariance matrix that are learned in different ways.
First, we consider the generic case, in which the mean and the covariance matrix are obtained through representations learned by means of a fully-connected neural network applied to the current measurement and the previous simulated state.
Second, we consider a recurrent neural network that is capable of capturing past information beyond the immediate previous simulated state.
Third, we make a case for the possibility of using neural network architectures that exploit additional data structure, if available.
In particular, we assume that the nonlinear dynamic system may be explained in terms of a distributed plant, and leverage graph neural networks to exploit this structure.
The fourth and last method is a non-parametric one, in which we learn an arbitrary mapping between samples from a uniform distribution and the state simulation.

We ran several simulations studying the performance of the proposed method against the sampling distribution designed to minimize particle degeneracy.
Overall, while the learned sampling distributions generally outperform the designed, minimum-degeneracy one, it is the RNN-based ones that consistently does so in a wide range of scenarios, with stable performance for a wide range of sweeping problem hyperparameters.
Note, however, that in the case of complex nonlinear systems, the other methods can provide better estimations, as seen in the case of the SIR system.
The FCNN and the arbitrary distribution methods are more flexible, allowing them to learn sampling distributions that adjust better to each instant of time and thus lead to better results.

This paper presents a first approach to learning sampling distributions in an unsupervised manner as opposed to designing them.
There are many directions for improvement of the methods presented herein.
First of all, more complex architectures can be considered, which may be better tailored for different specific nonlinear dynamical systems.
Second, the training can be improved by including regularization techniques such as dropout or penalties.
These may be particularly useful when the dimension of the systems is large.
Third, normalizing flows arise as an interesting direction to pursue.
The fourth method presented here is a very elementary normalizing flow.
Using more complex conditioning models to ensure the invertibility of the neural network may actually lead to increased  
representation capability and more sophisticated distributions.


%
%
%
%




\bibliographystyle{bibFiles/IEEEtranD}
\bibliography{bibFiles/myIEEEabrv,bibFiles/biblioPF}


%
%
%
%
%
%
%

\end{document}